\documentclass[aps,twocolumn,prb,superscriptaddress,showpacs,floatfix]{revtex4-1}
\pdfoutput=1
\usepackage{graphicx,xcolor}
\usepackage{dcolumn}
\usepackage{amsmath,amsfonts,amssymb,amsxtra}
\usepackage{mathrsfs}
\usepackage{bm}
\usepackage{times}
\usepackage{color,soul}

\def\MgB2{MgB$_{2}$}
\def\cm-1{cm$^{-1}$\,}
\def\cmT-1{cm$^{-1}$/T\,}
\def\E2g{$E_{\mbox{2g}}$}

\def\A1g{$A_{\mbox{1g}}$}
\def\2DS{$2\Delta_{S}^{E}$}

\def\2DA{$2\Delta^{A}$}
\def\D0{$2\Delta_{0}$}

\def\D6h{D$_{6h}$}

\begin{document}

\title{Critical Quadrupole Fluctuations and Collective Modes in Iron Pnictide Superconductors}
%
%
\author{V.~K.~Thorsm{\o}lle}
\email{vthorsmolle@physics.ucsd.edu}
\affiliation{Department of Physics and Astronomy, Rutgers, The State University of New Jersey, Piscataway, New Jersey 08854, USA}
\affiliation{Boston University, Department of Physics, Boston, MA 02215, USA}
\affiliation{University of California San Diego, Department of Physics, La Jolla, CA 92093, USA}
\author{M.~Khodas}
\affiliation{Department of Physics and Astronomy, University of Iowa, Iowa City, Iowa 52242, USA}
\affiliation{Racah Institute of Physics, The Hebrew University, Jerusalem 91904, Israel}
\author{Z.~P.~Yin}
\affiliation{Department of Physics and Astronomy, Rutgers, The State University of New Jersey, Piscataway, New Jersey 08854, USA}
\author{Chenglin~Zhang}
\affiliation{Department of Physics and Astronomy, Rice University, Houston, Texas 77005, USA}
\affiliation{Department of Physics, University of Tennessee, Knoxville, Tennessee 37996, USA}
\affiliation{Huazhong University of Science and Technology, National Pulse High Magnetic Field Center, Wuhan, 430074, China}
\author{S.~V.~Carr}
\affiliation{Department of Physics and Astronomy, Rice University, Houston, Texas 77005, USA}
\author{Pengcheng~Dai}
\affiliation{Department of Physics and Astronomy, Rice University, Houston, Texas 77005, USA}
\author{G.~Blumberg}
\email{girsh@physics.rutgers.edu}
\affiliation{Department of Physics and Astronomy, Rutgers, The State University of New Jersey, Piscataway, New Jersey 08854, USA}
\affiliation{National Institute of Chemical Physics and Biophysics, 12618 Tallinn, Estonia}
\date{\today}

\begin{abstract}
The multiband nature of iron pnictides gives rise to a rich temperature-doping phase diagram of competing orders and a plethora of collective phenomena. At low dopings, the tetragonal-to-orthorhombic structural transition is closely followed by a spin density wave transition both being in close proximity to the superconducting phase. A key question is the nature of high-$T_c$ superconductivity and its relation to orbital ordering and magnetism. Here we study the NaFe$_{1-x}$Co$_{x}$As superconductor using polarization-resolved Raman spectroscopy. The Raman susceptibility displays critical enhancement of non-symmetric charge fluctuations across the entire phase diagram which are precursors to a $d$-wave Pomeranchuk instability at temperature $\theta(\mbox{x})$. The charge fluctuations are interpreted in terms of quadrupole inter-orbital excitations in which the electron and hole Fermi surfaces breathe in-phase. Below $T_c$, the critical fluctuations acquire coherence and undergo a metamorphosis into a coherent ingap mode of extraordinary strength.
\end{abstract}

 \maketitle

\section{Introduction}
An important aim in the study of iron-based superconductors is to elucidate the nature of the superconducting state and its relation to adjacent phases \cite{Paglione2010,Wang2011,Chubukov2012}. Most FeAs compounds share a common phase diagram in which the underdoped region is marked by a tetragonal-to-orthorhombic structural transition at $T_S$ followed by a magnetic ordering transition at $T_{SDW}$ of collinear spin stripes which either precedes or coincides with $T_S$ \cite{Zhao2008,Li2009}. On introducing dopant atoms, superconductivity emerges with a transition temperature $T_c$ of tens of degrees \cite{Paglione2010}. The driving force behind the structural transition is widely debated with main proposals of either spin \cite{Xu2008,Fang2008,Fernandes2012,Fernandes2014} or ferro-orbital \cite{Kruger2009,Lv2011,Kontani2011,Lee2009c,Chen2010,Onari2012,Lee2012,Stanev2013,Yamase2013aa} nematic ordering. In the spin-nematic scenario the structural transition at $T_S$ is driven by magnetic fluctuations which breaks fourfold rotational ($C_4$) lattice symmetry \cite{Fernandes2012,Fernandes2014}. The latter induces a sharp increase of the spin correlation length for one spin stripe orientation and a decrease of the other. In the orbital-nematic scenario, $C_4$ symmetry is broken by ferro-orbital ordering in which strong inter-orbital interactions lead to inequivalent occupation of the $d_{xz}$ and $d_{yz}$ Fe orbitals.

An enhancement of spin susceptibility is observed in INS or NMR pnictide data upon approaching the SDW transition \cite{Ma2011,Lu2014}. However, at higher dopings away from the SDW phase, this enhancement is rapidly suppressed \cite{Lu2014,Nakai2013,Zhang2013}. Hence, while the close proximity to magnetic order naturally favors spin fluctuations as a candidate in providing the glue for Cooper pairs \cite{Chubukov2012}, suppressed spin fluctuations appear to be insufficient in explaining the whole temperature-doping ($T$$-$x) phase diagram \cite{Yamase2013b}. NMR measurements of the relaxation rate $1/T_1T$ in FeSe, which has no SDW transition, revealed that spin fluctuations only emerge below $T_S$ and the nematic order was argued to be driven by orbital degrees of freedom \cite{Bohmer2015a,Baek2015}.

\begin{figure*}[htpb]
  \vspace*{-10pt}\centering
  \includegraphics[width=\textwidth]{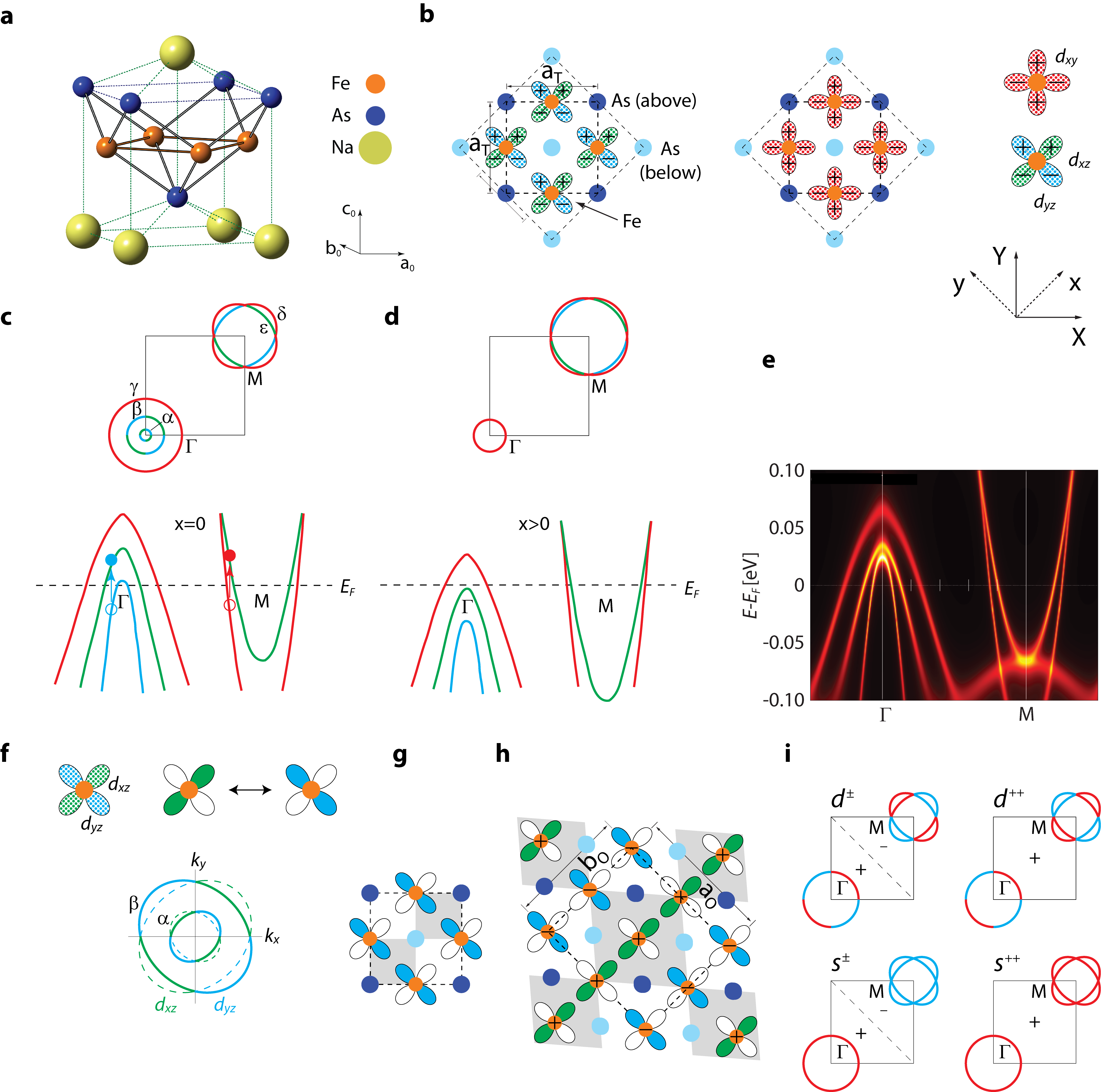}
  \caption{NaFe$_{1-x}$Co$_{x}$As crystal and electronic structure, and $XY$$-$quadrupole mode. (a) Crystal structure of NaFeAs in the tetragonal phase. (b) Top view of FeAs layer in the tetragonal phase shown with $d_{xz}$$-$$d_{yz}$ orbitals (left) and $d_{xy}$ orbitals (right). Dashed lines represent the two/four-Fe unit cell in the tetragonal/orthorhombic phase. (c and d) The effect of Co-doping is illustrated on the schematic Fermi surfaces (FS) for NaFe$_{1-x}$Co$_{x}$As in the tetragonal nonmagnetic BZ for doping $x$=0 (c) and $x$$>$0 (d). Below is shown a band-dispersion cut along the $\Gamma$$-$M high-symmetry line. $d_{xy}$, $d_{xz}$ and $d_{yz}$ orbitals are shown with respectively red, blue and green colors. The hole-like pockets $\alpha$, $\beta$ and $\gamma$ surround the $\Gamma$ point, and the electron-like pockets $\varepsilon$/$\delta$ surround the M point. (e) Momentum- and frequency-resolved spectra $A(\mbox{\bf{k}},\omega)$ along the $\Gamma$$-$M high-symmetry line calculated by first-principle calculations including spin-orbit coupling (See Appendix D). (f) Pomeranchuk fluctuations in $B_{2g}$ symmetry which is sustained by charge transfers between degenerate $d_{xz}$ and $d_{yz}$ Fe-orbitals. (See text and Appendix E). (g) Monoclinic 2-Fe unit cell in the Pomeranchuk phase. (h) Quadrupole groundstate in the orthorhombic phase with orthorhombic structural distortion, doubled unit cell and two neighboring stripes having different orbital occupation. The pluses and minuses indicate a buckling-like modulation effect along the $c$-axis. (i) Phase of the superconducting OPs for the $\gamma$ band at the $\Gamma$-point and the $\delta$/$\varepsilon$ bands at the M-point for $s^{++}$, $d^{++}$, $s^{\pm}$, and $d^{\pm}$ symmetry. Different colors indicate opposite sign of the gap function.
   \vspace*{0pt}}\label{Fig: FeAs}
\end{figure*}

In elastic strain measurements of Co- and K-doped BaFe$_2$As$_2$ and FeSe, the shear modulus $C_{66}$ softens on cooling and 1/$C_{66}$ follows a Curie-Weiss-like behavior which is interrupted at $T_S({\mbox{x}}$). Elastoresistivity measurements display a similar behavior \cite{Chu2012,Kuo2015}. The Weiss-temperature, which we define as $\theta(\mbox{x})$ is observed to increase towards zero doping \cite{Bohmer2014,Bohmer2015Ar,Yoshizawa2012,Goto2011,Chu2012}. $T_S-\theta$ was related to the contribution of the lattice to the electronic nematic fluctuations in Refs.~\onlinecite{Bohmer2014,Bohmer2015Ar}. $T_S-\theta$ correlates with the downshift of the mean field transition temperature $\theta(\mbox{x})$ \cite{Chu2012}. These two temperatures being noticeably different and a non-vanishing 1/$C_{66}$ at $T_S$ leaves the origin of the transition $\theta(\mbox{x})$ as an open question. The 1/$C_{66}$ temperature dependence was attributed to the electric quadrupole fluctuations due to the 3$d$ inter-orbital fluctuations in Refs.~\onlinecite{Yoshizawa2012,Goto2011}. This conjecture is supported by detection of Fe-quadrupole orbital fluctuations by electron diffraction measurements \cite{Ma2014}. At present it is unsettled whether the $\theta(\mbox{x})$-line \cite{Bohmer2015Ar,Kontani2014} is associated with the structural instability or is a separate instability of a different nature that breaks $C_4$ symmetry.

The nematic theories and analysis of experimental data are generally based on the assumption that $C_4$ symmetry is broken at $T_S$ while translational symmetry is broken at $T_{SDW}$ \cite{Fernandes2012,Fernandes2014,Bohmer2014,Bohmer2015Ar,Yoshizawa2012,Goto2011,Chu2012}. Most pnictides have $T_S$ and $T_{SDW}$ near-conjoint in the low-doping regime \cite{Cano2010} including the heavily studied 122-family, i.e. Co- or K-doped BaFe$_2$As$_2$. However, NaFe$_{1-x}$Co$_{x}$As, which is a 111-system, has $T_S$ and $T_{SDW}$ separated by more than 10~K and presents a better suited material in which the nature of the structural and SDW transitions can be studied separately \cite{Steckel2015,Deng2015,Wright2012,Tan2013}.

Temperature-dependent X-ray powder diffraction studies of NaFe$_{1-x}$Co$_{x}$As find that at $T_S$ the high-temperature tetragonal P4/nmm structure transforms into the orthorhombic Cmma structure with the orthorhombic distortion $\delta$=a$_O$-b$_O$ emerging smoothly upon cooling \cite{Wright2012}. Here, a$_O$ and b$_O$ are the lattice parameters of the orthorhombic unit cell, a$_O$=$\sqrt{2}$a$_T$+$\delta /2$ and b$_O$=$\sqrt{2}$a$_T$-$\delta /2$ (Fig.~\ref{Fig: FeAs}(h)), and a$_T$ of the tetragonal unit cell (Fig.~\ref{Fig: FeAs}(b)). Neutron diffraction and muon spin rotation data confirms that the lattice distortion starts above $T_{SDW}(\mbox{x})$ \cite{Parker2010,Li2009}, and an ARPES study reports Brillouin zone (BZ) folding and doubling of the unit cell at $T_S(\mbox{x})$ \cite{He2010}. The orthorhombic OP is established at $T_S$ and development of SDW long-range order is established at $T_{SDW}$ \cite{Park2012,Li2009}. The smooth continuous OP implies the occurrence of a single structural instability which sets in at $T_S(\mbox{x})$. The structural transition appears to be subtle with the volume of the lattice changing only marginally \cite{Li2009} while both transitions display anomalies in resistivity measurements \cite{Steckel2015,Wang2013}.
Specific heat studies reveal anomalies at $T_S$ and $T_{SDW}$ which are characteristic of second-order phase transitions \cite{Steckel2015,Wang2012a,Chen2009}.
The spin-nematic scenario predicts a jump in the magnetic correlation length at $T_S(\mbox{x})$ and formation of a pseudogap which has been refuted by INS measurements where the $T_{SDW}(\mbox{x})$ and $T_S(\mbox{x})$ transitions appear to be decoupled \cite{Park2012}. This may be in contrast to that observed in the 122 systems (Ba,Ca)Fe$_2$As$_2$ \cite{Kim2011,Goldman2008} implying the spin-nematic scenario is more applicable to 122 systems than to NaFe$_{1-x}$Co$_{x}$As. An alternative picture to the spin-nematic model is where critical ferroquadrupoles trigger the orthorhombic structure transition which involves a ferro-orbital density wave at $T_S(\mbox{x})$ \cite{Kontani2011}.

The proximity of the structural and superconducting phase transitions is universal which makes it necessary to investigate both instabilities in one setting. So far, no clear consensus has been reached on the symmetry of the superconducting OP. Theories building on spin fluctuations favor unconventional $s^{\pm}$ pairing in which the superconducting OP changes sign between electron- and hole-like FSs \cite{Chubukov2012,Mazin2008}. Yet, other theories embrace orbital fluctuations building on superconductivity with $s^{++}$-pairing in which there is no sign change \cite{Kontani2010}. Recently, orbital antiphase $s^{\pm}$ has been proposed in which the pairing function of the Fe $d_{xy}$ orbital has opposite sign to the $d_{xz}$ and $d_{yz}$ orbitals \cite{Yin2014}, as well as orbital triplet pairing \cite{Ong2013}. The type of doping leading to superconductivity can either have a nodeless ($s$-wave) or a nodal ($d$-wave) OP (Fig.~\ref{Fig: FeAs}(i)). Electron- or hole-doping BaFe$_2$As$_2$ with respectively Co or K leads to a nodeless OP, except at high hole-dopings where a switch to a nodal OP occurs. In contrast, isovalent substitution with P yields a nodal OP \cite{Shibauchi2014}.

A long-standing issue that remains unresolved in many classes of unconventional superconductors, including cuprates \cite{Moon2012,Moon2012a}, heavy fermions \cite{Wolfle2011,Lohneysen2007}, and iron pnictides is whether a quantum critical point (QCP) lies beneath the superconducting dome \cite{Sachdev2012,Shibauchi2014}. The quantum criticality related to the anti-ferromagnetic QCP was extensively studied within the spin fermion model\cite{Abanov2003,Metlitski2010,Moon2012,Moon2012a}. In this model the critical fluctuations related to the QCP were shown to affect the properties far into the normal state. Hence, the existence and detection of a QCP may offer an understanding of the origin of unconventional superconductivity and its coexistence with either magnetic or exotic phases. It has recently been demonstrated in theoretical studies that Cooper pairing is enhanced in the vicinity of a nematic QCP \cite{Maier2014,Lederer2015}. Experimentally, elastic anomalies of the $C_{66}$ shear modulus observed near a QCP in Ba(Fe$_{1-x}$Co$_{x}$)$_2$ As$_2$ suggests the involvement of ferro-quadrupole fluctuations \cite{Yoshizawa2012,Goto2011,Kontani2011}.
A second-order quantum phase transition lying beneath the superconducting dome has been reported in BaFe$_2$(As$_{1-x}$P$_{x}$)$_2$ by measurements of the London penetration depth \cite{Hashimoto2012}. However, a study using NMR, X-rays and neutrons finds no signatures of a QCP \cite{Hu2015} raising questions as to its origin. Identification of the charge multi-polar collective excitations and their symmetry associated with the nematic QCP is essential for understanding superconductivity and competing phases to which Raman spectroscopy is the most suitable probe \cite{Klein1984,Klein2010,Klein2009,Devereaux2007,Wai2013,Gallais2015A,Hinojosa2015}.

We use polarization-resolved electronic Raman spectroscopy to study the charge dynamics of the multiband NaFe$_{1-x}$Co$_{x}$As superconductors characterized by partially filled 3$d$-orbitals. We demonstrate that charge transfers between the degenerate $d_{xz}$ and $d_{yz}$ orbitals lead to collective intra-orbital quadrupole charge fluctuations in the normal and superconducting state. We find that the entire tetragonal phase is governed by the emergence of strong overdamped orbital quadrupole fluctuations which upon cooling display critical enhancement. These critical fluctuations foretell an approaching subleading second order phase transition with broken $C_4$ symmetry and an orbitally-ordered state. In the low doping region, the formation of this phase is intervened by the structural transition and becomes subleading. Below $T_c$, the fluctuations acquire coherence and undergo a metamorphosis into a sharp ingap mode of extraordinary strength.

In Section II we introduce the Raman experiments including sample preparations and the Raman probe. In Section III we give an overview of the NaFe$_{1-x}$Co$_{x}$As Raman data and establish the $T$$-$x phase diagram of the static Raman susceptibility. In Section IV we compare the static Raman susceptibility to a two-component fit of the NMR relaxation rate. In Section V we present and analyse the Raman data in more details and discuss it in terms of critical quadrupole fluctuations and the Pomeranchuk instability. In Section VI we discuss a possible density wave state below the structural transition. In Section VII we present Raman data in the superconducting state which entails discussions of ingap collective modes and their connection to critical quadrupole fluctuations in the normal state. In Section VIII we present the Bardasis-Schrieffer mode and its interplay with the ingap exciton mode in the particle-hole channel. In Section IX we discuss a quantum critical point inside the superconducting dome in terms of the Pomeranchuk instability and the interplay of the Bardasis-Schrieffer mode with the strong ingap collective mode. In Section X we present the main conclusion of the $d$-wave Pomeranchuk quadrupole fluctuations and their relation to the ingap collective mode of extraordinary strength. In The Appendices we present Appendix A: Analysis of Raman Spectra; Appendix B: Coupling of Pomeranchuk Fluctuations to the Raman Probe; Appendix C: Relaxational Mode Fitting Procedure; Appendix D: First-Principle Band Structure Calculations; and Appendix E: Symmetry Modes in Momentum Space.

\section{Methods}


\subsection{Sample Preparation}

NaFe$_{1-x}$Co$_{x}$As single crystals were grown by the self-flux method as described in Ref.~\onlinecite{Tanatar2012}. The volume fractions of bulk superconductivity for compounds with a doping range between 0.015 and 0.06, measured with a Quantum Design SQUID magnetometer, were larger than 80$\%$. $T_S$, $T_{SDW}$ and $T_c$ versus doping were reported in Ref.~\onlinecite{Tan2013} and are shown in the $T$$-$x phase diagram, Fig.~\ref{Fig: PhaseDiagram}(a). The superconducting gap values 2$\Delta$ determined by ARPES in Refs.~\onlinecite{Ge2013,Liu2011} are indicated by vertical dashed lines in Fig.~\ref{Fig: SCModes}. Figure~\ref{Fig: SM_Blue}(a) shows 2$\Delta$ determined by Raman. The samples were vetted for the highest quality surfaces and were handled in a protective argon atmosphere in a glovebox, where they were packed into sealed glass containers with a protective argon atmosphere. Upon preparing to do the Raman measurements the sample was unpacked inside a nitrogen-filled protective glovebag sealed to the entrance of the cryostat. The crystal was then cleaved and positioned in the continuous flow optical cryostat.


\subsection{Experimental Methods}

All Raman scattering measurements were performed in a quasi-back scattering geometry along the crystal $c$-axis and excited with a Kr$^+$ laser line. We used a laser excitation energy of $\omega_L$=2.6~eV, except for investigations of the ingap collective modes shown in Figs.~\ref{Fig: SCModes_Red},\ref{Fig: CollectiveModes}(a) where $\omega_L$=1.93~eV was also used. The incident laser power was less than 12~mW focused to a 50$\times$100~$\mu$m$^2$ spot on the $ab$-surface. In the superconducting state the power was reduced to less than 2~mW. For $\omega_L$=2.6~eV, being close to resonant condition, the lowest temperature was $\simeq$5~K. The lower excitation energy of $\omega_L$=1.93~eV being pre-resonant and at $\simeq$3~K allowed us to observe both the $\omega^{p\mbox{-}p}_{B_{2g}}$ as well as the $\omega^{p\mbox{-}h}_{B_{2g}}$ excitons at finite frequencies.
The spectra of the collected scattered light were measured by a triple-stage Raman spectrometer designed for high-straylight rejection and throughput equipped with a liquid nitrogen-cooled charge-coupled detector.

The Raman spectra were corrected for the spectral response of the spectrometer and detector in obtaining the Raman scattering intensity, $I_{e^Ie^S}(\omega) = (1 + n)\chi''(\omega) + L(\omega)$. Here, $L(\omega)$ is a small luminescence background and $\mbox{\bf e}^{I}$ and $\mbox{\bf e}^{S}$ the polarization vectors for the incident and scattered photons for a given scattering geometry with respect to the unit cell (Fig.~\ref{Fig: FeAs}(b)). The recorded Raman intensity was background subtracted with a near-linear line and a constant determined for each polarization geometry (See Appendix A).

In obtaining the static Raman susceptibility $\chi^{XY}_0(T,\mbox{x})$ in the $B_{2g}$ symmetry channel shown in Fig.~\ref{Fig: PhaseDiagram} we performed a K-K transformation of the $\chi''_{XY}(\omega,T,\mbox{x})$ data shown in Figs.~\ref{Fig: RamanData2}(d-f),\ref{Fig: RamanData3}(a-e). For a given doping x, the $\chi''_{XY}(\omega)$ spectra for each temperature was first divided by $\omega$ to obtain $\chi''_{XY}(\omega)/\omega$. The lower frequency cutoff is $\simeq$20~cm$^{-1}$ and $\chi''_{XY}(\omega)/\omega$ was therefore extended to zero frequency with a phenomenological even function which fits well to the data. The static Raman susceptibility was then calculated from the Kramers-Kronig relation,
\begin{equation}
 \chi_{XY}'(0)=\chi^{XY}_0=\frac{2}{\pi} P\int^{\infty}_0\frac{\chi''_{XY}(\omega)}{\omega}d\omega
\label{KramersKronig}
\end{equation}
at zero frequency. The integration was performed up to the highest measured frequency $\simeq$750~cm$^{-1}$ at which point $\chi''_{XY}(\omega)/\omega$ was near zero.


\subsection{The Raman Probe}

The Raman response function is sensitive to charge density fluctuations driven by the incident and scattered photon fields. For a given scattering geometry with polarization vectors $\mbox{\bf e}^{I}$ and $\mbox{\bf e}^{S}$ for the incident and scattered photons, the Raman susceptibility is given by,

\begin{equation}
\chi_{I,S}(\omega)\propto -i\int_{0}^{\infty} e^{i\omega t}
\left<[\widetilde{\rho}^{I,S}(t),\widetilde{\rho}^{I,S}(0)]\right> \, dt.
\label{response}
\end{equation}

The symmetrized Raman tensor $\chi_{I,S}(\omega)$ for the different scattering geometries can be classified by the irreducible representations for the crystallographic point group \cite{Ovander1960}. The symmetry channels accessible by Raman scattering transform $A_{1g}$, $A_{2g}$, $B_{1g}$, $B_{2g}$ and $E_g$ irreproducible representation of $D_{4h}$ point group (above $T_S$) and as $A_g$, $B_{1g}$, $B_{2g}$ and $B_{3g}$ for $D_{2h}$ (below $T_S$). Below $T_S$, $A_{1g}$ and $B_{2g}$ becomes $A_{g}$. Using circularly polarized light we confirmed that the contribution
from the $A_{2g}$ symmetry channel can be neglected. The scattering geometry is referenced to the $X$$-$$Y$ coordinate system of the crystallographic (As-As) unit cell depicted in Fig.~\ref{Fig: FeAs}(b). The incident and scattered photon fields cross polarized along the $a$- and $b$-directions of the two-Fe unit cell yields $\chi_{XY}(\omega)$ susceptibility. For NaFe$_{1-x}$Co$_{x}$As with $D_{4h}$ point group symmetry in the tetragonal phase, $\chi_{XY}(\omega)$ probes excitations in $B_{2g}$ symmetry. The cross polarized photon fields rotated by 45$^{\circ}$ yields $\chi_{xy}(\omega)$ or $B_{1g}$ susceptibility. $\chi_{A_{1g}}(\omega)$ can be obtained in two steps: first, by aligning both photon fields along one axis, and then by obtaining the $xy$ susceptibility. $\chi_{A_{1g}}(\omega)$ is given by $\chi_{XX}(\omega)$$-$$\chi_{xy}(\omega)$. The space group in the tetragonal and orthorhombic phase is respectively $P4/nmm$ with point group $D_{4h}$ and $Cmma$ with point group $D_{2h}$ \cite{Li2009}. Entering the orthorhombic phase from the tetragonal phase is associated with broken symmetry operators which includes $C_4$ rotations and mirror planes of the tetragonal phase with $D_{4h}$ point group symmetry. The point group symmetry for the orthorhombic phase is $D_{2h}$ in which both $B_{2g}$ and $A_{1g}$ symmetry of the tetragonal phase conforms to $A_g$ symmetry.

\section{Overview of NaFe$_{1-x}$Co$_{x}$As Raman Data}

\begin{figure*}[htpb]
  \vspace*{-10pt}\centering
  \includegraphics[width=\textwidth]{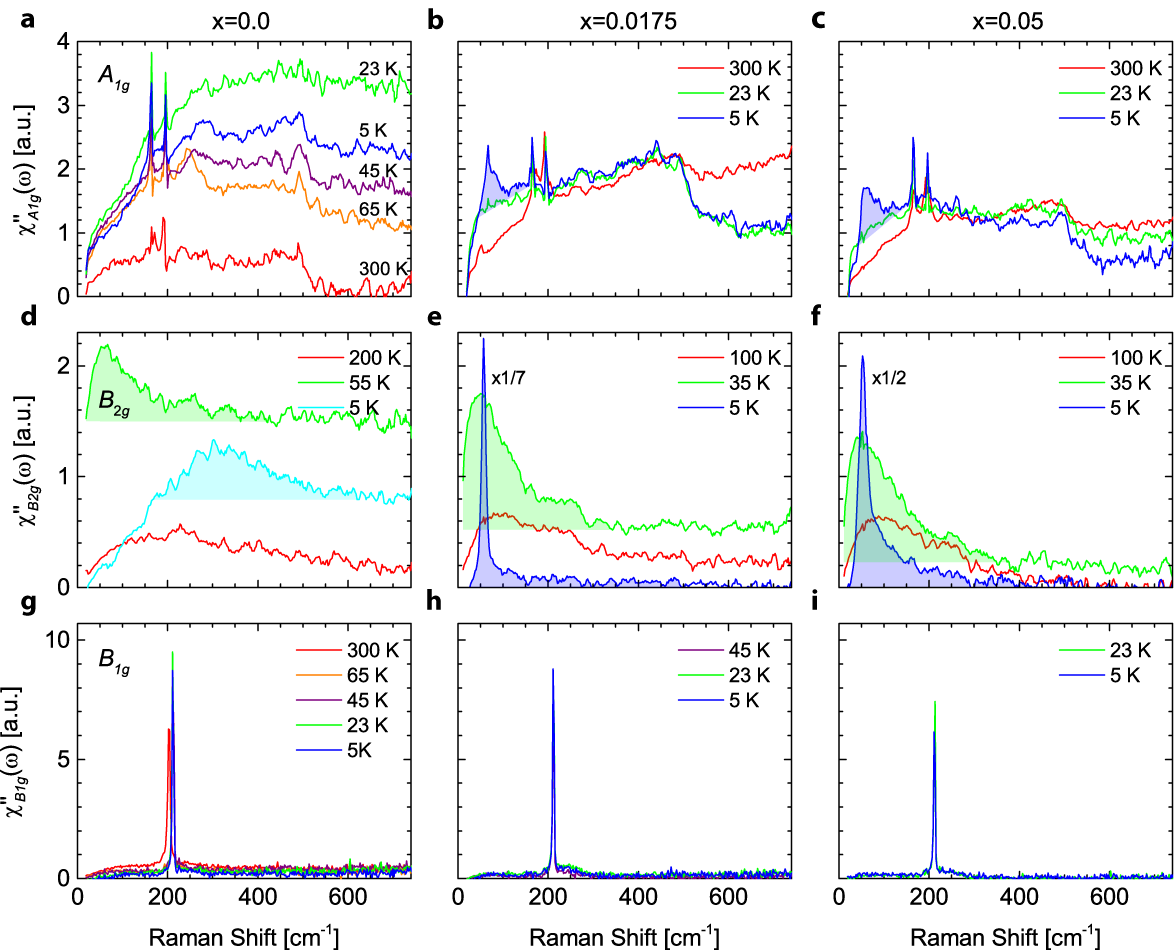}
  \caption{Raman susceptibility $\chi''(\omega)$ in the $A_{1g}$, $B_{2g}$ and $B_{1g}$ symmetry channels at representative temperature and dopings. (a to c) $\chi''_{A_{1g}}(\omega)$ showing superconducting features highlighted with blue shading below $\simeq$200~cm$^{-1}$ (x=0.0175, x=0.05). (d to f) $\chi''_{B_{2g}}(\omega)$ presenting a quasielastic scattering relaxational mode above $T_S(\mbox{x})$ and $T_c(\mbox{x})$ highlighted with green shading, a density wave suppression and coherence peak highlighted with light blue shading below $T_S(\mbox{x})$ (x=0, 5~K), and a low-temperature collective resonance highlighted with blue shading (x=0.0175, x=0.05, 5~K). (g to i) $\chi''_{B_{1g}}(\omega)$ featuring mainly a $B_{1g}$ phonon.
  \vspace*{0pt}}\label{Fig: RamanData}
\end{figure*}

The temperature or doping dependent electronic Raman susceptibility $\chi''({\omega},T,\mbox{x})$ reveals the dynamics of collective excitations and provides an unambiguous identification of their symmetry \cite{Klein1984,Klein2010,Klein2009,Devereaux2007}.
The symmetrized Raman tensor for the different scattering geometries can be classified by the irreducible representations for the crystallographic point group \cite{Ovander1960}. The symmetry channels accessible by Raman scattering are $A_{1g}$, $A_{2g}$, $B_{1g}$, $B_{2g}$ and $E_g$ for pnictides with a tetragonal 2-Fe unit cell, i.e. for NaFe$_{1-x}$Co$_{x}$As (above $T_S$).

\begin{figure*}[!t]
  \vspace*{-10pt}\centering
  \includegraphics[width=1.0\textwidth]{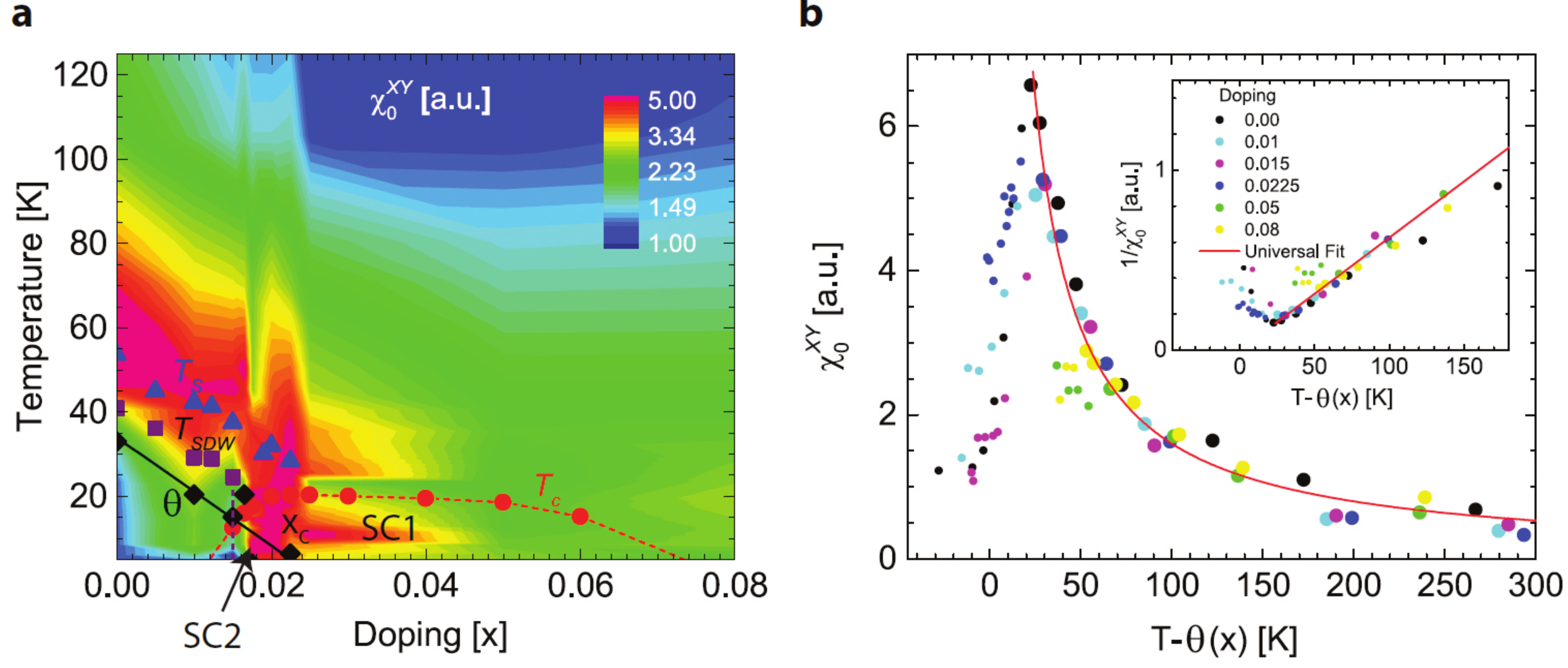}
  \caption{Static Raman susceptibility $\chi^{XY}_0(T,\mbox{x})$ in the $B_{2g}$ symmetry channel. (a) Evolution of $\chi^{XY}_0(T,\mbox{x})=2/\pi\int_0^{\infty}(\chi''_{XY}(\omega)/\omega)d\omega$ as a function of temperature and doping. The structural transition $T_S(\mbox{x})$, the magnetic transition $T_{SDW}(\mbox{x})$, and the superconducting transition temperature $T_c(\mbox{x})$ (from Ref.~\onlinecite{Tan2013}) are indicated by blue triangles, purple squares and red circles respectively. $\theta(\mbox{x})$ is the mean field transition temperature associated with the critical behavior of $\chi^{XY}_0(T,\mbox{x})$. (b) $\chi^{XY}_0(T,\mbox{x})$ is shown with a universal fit to $A/(T-\theta(\mbox{x}))$ where the temperature axis for each doping x is shifted by $\theta(\mbox{x})$. The inset shows the inverse of $\chi^{XY}_0(T,\mbox{x})$ versus temperature $T-\theta(\mbox{x})$ with a fit to a universal straight line.
  \vspace*{0pt}}\label{Fig: PhaseDiagram}
\end{figure*}

In Fig.~\ref{Fig: RamanData} we show Raman susceptibility
$\chi''(\omega)$ at representative temperatures and dopings for
the $A_{1g}$, $B_{2g}$ and $B_{1g}$ symmetry channels to point out
important features in relation to the tetragonal, orthorhombic, SDW
and superconducting phases of the $T$$-$x phase diagram
(Fig.~\ref{Fig: PhaseDiagram}(a)) which will be discussed in depth
below. Most of these features are reflected in the $\chi''_{B_{2g}}(\omega)$
response, while $\chi''_{A_{1g}}(\omega)$ contains important characteristics
of superconducting nature, and $\chi''_{B_{1g}}(\omega)$ mainly features a
$B_{1g}$ phonon. The detailed temperature and doping dependence is
shown in Figs.~\ref{Fig: RamanData2}-\ref{Fig: SCModes_Red}.


Using Kramers-Kronig (K-K) transformation, we calculate the real part of $\chi_{XY}(\omega,T,\mbox{x})$ at $\omega$=0, the static
Raman susceptibility $\chi_0^{XY}(T,\mbox{x})$, for $B_{2g}$ symmetry.
Figure~\ref{Fig: PhaseDiagram}(a) shows $\chi^{XY}_0(T,\mbox{x})$ in a
$T-$x phase diagram where $T_S(\mbox{x})$, $T_{SDW}(\mbox{x})$, and
$T_c(\mbox{x})$ obtained by transport measurements \cite{Tan2013} are
superimposed on top. The enhancement of $\chi^{XY}_0(T,\mbox{x})$
with cooling, observed for all x, starts from high temperatures and
culminates in a maximum at the structural transition $T_S(\mbox{x})$
or at a smaller maximum before the $T_c(\mbox{x})-$line for higher
dopings. $\chi^{XY}_0(T,\mbox{x})$ is suppressed below the structural
transition $T_S(\mbox{x})$ \cite{Note3}. Figure~\ref{Fig: PhaseDiagram}(b) shows $\chi_0^{XY}(T,\mbox{x})$ with a universal fit to $A/(T-\theta(\mbox{x}))$ where the temperature axis for each doping x is shifted by $\theta(\mbox{x})$. The inset shows the inverse of $\chi_0^{XY}(T,\mbox{x})$ versus temperature $T$$-$$\theta(\mbox{x})$ with a fit to a universal straight line.


The two sharp modes in $\chi''_{A_{1g}}(\omega)$ at $\simeq$164 and
$\simeq$195~cm$^{-1}$, and in $\chi''_{B_{1g}}(\omega)$ at
$\simeq$211~cm$^{-1}$ observed in the spectra for all dopings and
temperatures are phonon excitations (Figs.~\ref{Fig: RamanData}(g,h,i)),
as they are expected for the 111-family crystallographic structure
\cite{Um2012,Um2014}. The frequencies of these phonons increase slightly
with cooling, typical of anharmonic behavior, and do not display any
anomalies in self-energy upon crossing phase transition lines.


For low dopings, the $\chi''_{A_{1g}}(\omega)$ susceptibility displays an
overall enhancement of the spectra upon traversing the
high-temperature tetragonal phase to the orthorhombic and SDW phases
which maximizes at lower temperatures. For x$\gtrsim$0.0175, the most
important changes occur in the low-frequency region below
$\simeq$200~cm$^{-1}$ when crossing from the normal into the
superconducting state. Here $\chi''_{A_{1g}}(\omega)$ displays markedly
different dynamics above and below $T_c(\mbox{x})$, with featureless
spectra above $T_c(\mbox{x})$ and below, one or more superconducting
features in the range of $\simeq$70~cm$^{-1}$.



The $B_{2g}$ symmetry channel, $\chi''_{XY}(\omega)$ contains several
characteristics: (i) a broad peak extending to about 400~cm$^{-1}$,
indicated by green shading, which is dominating in the entire
tetragonal phase above the $T_S(\mbox{x})$ and $T_c(\mbox{x})$ lines;
(ii) a low-frequency suppression and coherence peak in the
orthorhombic phase, indicated by light blue shading; (iii) a sharp
resonance of extraordinary strength in the superconducting phase at $\simeq$$57$~cm$^{-1}$ (7.1~meV), indicated by blue shading;
(iv) a broad continuum which diminishes with doping (Fig.~\ref{Fig: RMFit}(a)).

\section{Static Raman Susceptibility and NMR $1/T_1T$ Relaxation Rate}

\begin{figure*}[htpb]
 \vspace*{-10pt}\centering
 \includegraphics[width=1.0\textwidth]{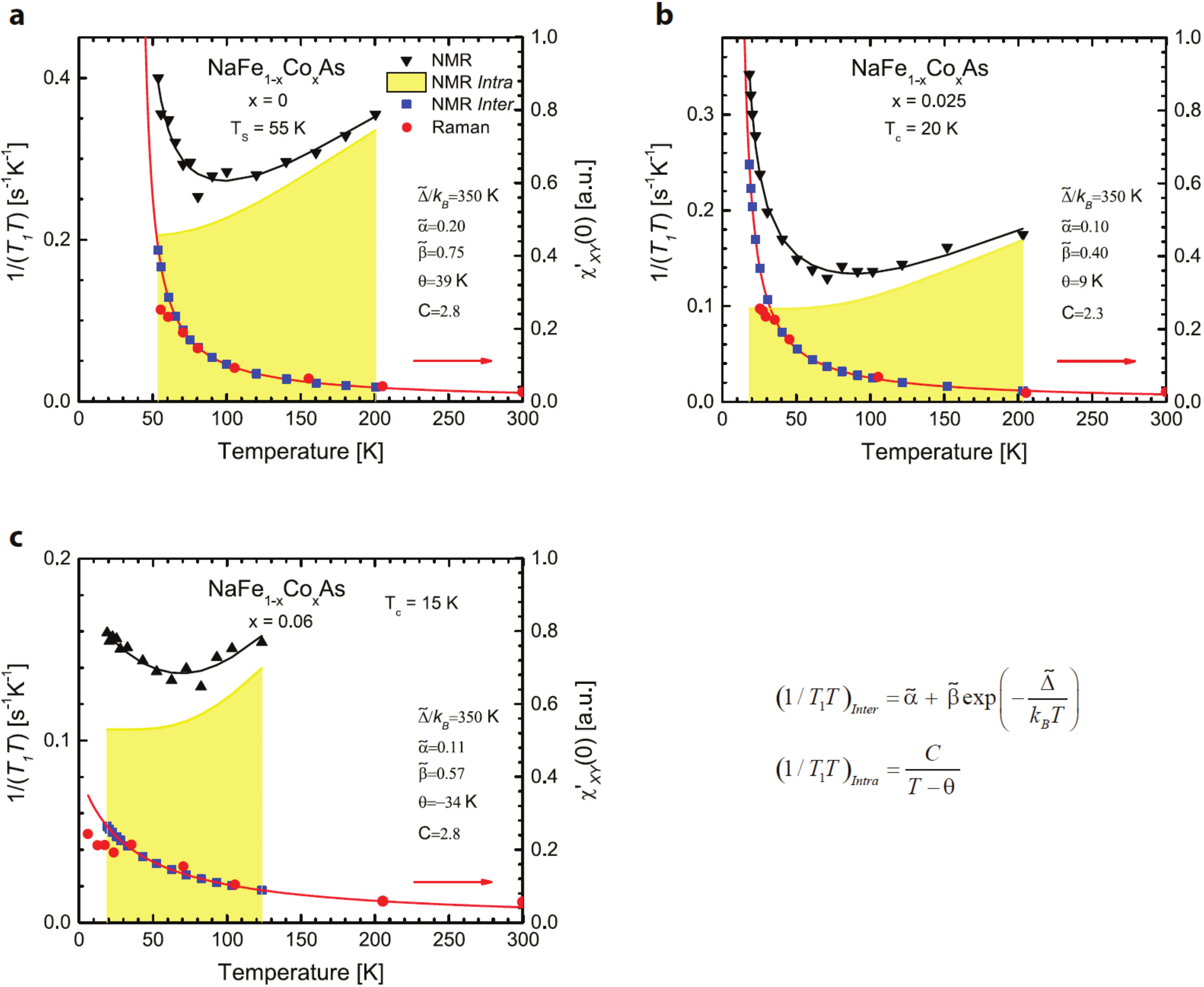}
 \caption{Two-component fit of the NMR relaxation rate $1/^{75}T_1T$ for As in NaFe$_{1-x}$Co$_x$As compared to the static Raman susceptibility $\chi'_{XY}(0,T)$. Temperature dependence of the NMR relaxation rate $1/^{75}T_1T$ for As (black triangles) compared to the static Raman susceptibility $\chi'_{XY}(0,T)$ (red circles). $1/^{75}T_1T$ is decomposed into two contributions, $1/T_1T=(1/T_1T)_{Intra}+(1/T_1T)_{Inter}$ where $(1/T_1T)_{Intra}=C/(T-\theta)$ (yellow shades) and $(1/T_1T)_{Inter}=\tilde{\alpha}+\tilde{\beta} exp(-\tilde{\Delta}/k_BT)$ (blue squares). $\chi'_{XY}(0,T)$ scales to $(1/T_1T)_{Intra}$ for all three dopings x=0, x=0.025 and x=0.06 and we attribute $\theta$ to correspond to the Pomeranchuk transition temperature $\theta$ described in the main text, and where the red line is the Curie-Weiss fit. For the fits of $1/T_1T$ we have used $\tilde{\Delta}/k_B$=350 or $\tilde{\Delta}$=240~cm$^{-1}$. This value of $\Delta$ corresponds to the minor mode which is present in $\chi''_{XY}(\omega,T)$ for all dopings and temperatures above $T_S(\mbox{x})$. (a) NMR data for x=0 is from Ref.~\onlinecite{Ma2011}. (b) NMR data for x=0.025 is from Ref.~\onlinecite{Oh2013}. (c) NMR data for x=0.06 is from Ref.~\onlinecite{Ji2013}.\vspace*{0pt}}\label{Fig: SM_NQR}
\end{figure*}

Figure~\ref{Fig: SM_NQR} displays the temperature dependence of the NMR relaxation rate $1/^{75}T_1T$ for As compared to the static Raman susceptibility $\chi'_{XY}(0,T)$. The NMR data for dopings x=0, 0.025 and 0.06 are from Refs.~\onlinecite{Ma2011,Oh2013,Ji2013}, respectively. The As nucleus has spin 3/2 and can relax into both an electronic spin or a charge quadrupole excitation\cite{Dioguardi2015}. The latter is described by nuclear quadrupole resonance (NQR). $1/^{75}T_1T$ is decomposed into two contributions, $1/T_1T=(1/T_1T)_{Intra}+(1/T_1T)_{Inter}$ where $(1/T_1T)_{Intra}=C/(T-\theta)$ and $(1/T_1T)_{Inter}=\tilde{\alpha}+\tilde{\beta} exp(-\tilde{\Delta}/k_BT)$ \cite{Ning2010}. In this model, the former is the Curie-Weiss law due to intraband relaxation and the latter is due to interband-like excitations in which the gap $\tilde{\Delta}$=240~cm$^{-1}$ is used. $\chi'_{XY}(0,T)$ scales to $(1/T_1T)_{Intra}$ for all three dopings x=0, x=0.025 and x=0.06 and we attribute $\theta$ to correspond to the Pomeranchuk transition temperature at $\theta(\mbox{x})$. The used value for $\tilde{\Delta}$ corresponds to the minor mode at 240~cm$^{-1}$ which is present in $\chi''_{XY}(\omega,T)$ for all dopings and temperatures above $T_S(\mbox{x})$. The self-consistency of the presented analysis of $1/^{75}T_1T$ and its correspondence to $\chi'_{XY}(0,T)$ suggests that $1/^{75}T_1T$ for NaFe$_{1-x}$Co$_x$As originates from quadrupole excitations and not spin relaxation. These orbital singlet excitations have $\Delta$L=2 and can be detected by Raman spectroscopy and NQR but not INS experiments. The same scaling analysis, including elastic probes, applied to the 122-family (Sr,Eu)Fe$_2$As$_2$ in Ref.~\onlinecite{Zhang2014} implies this role of quadrupoles may be a general feature of pnictide materials.

\section{Critical Quadrupole Fluctuations}

\begin{figure*}[htpb]
 \vspace*{-10pt}\centering
 \includegraphics[width=\textwidth]{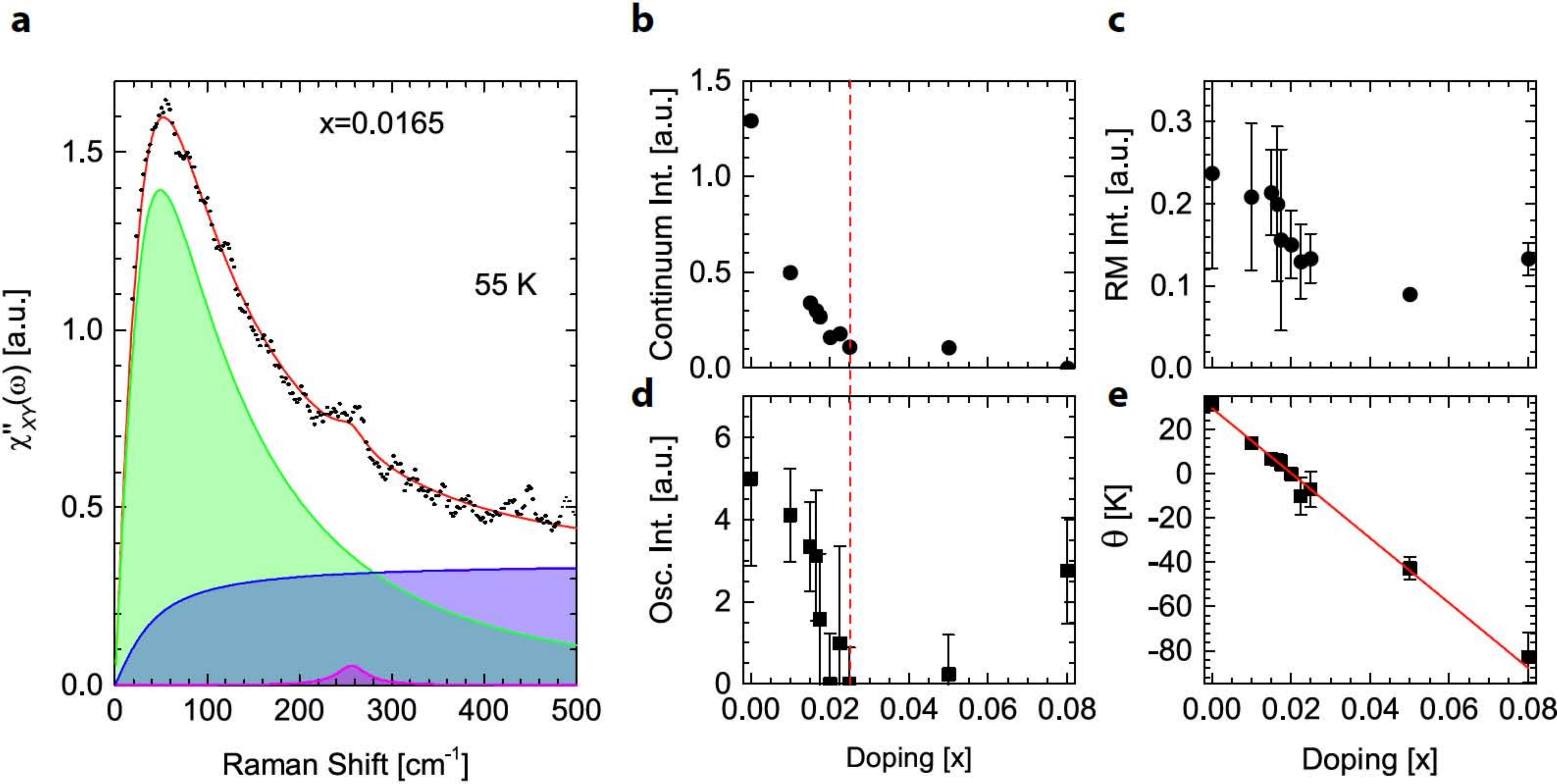}
 \caption{Decomposition of the Raman susceptibility $\chi''_{XY}(\omega,T,\mbox{x})$ in the $B_{2g}$ symmetry channel. (a) $\chi''_{XY}(\omega,T,\mbox{x})$ (black dots) for the representative doping x=0.0165 and temperature 55~K in the tetragonal phase. The red line is a fit to the data which are decomposed into three components: a continuum background (blue shading), a Lorentz oscillator (magenta shading), and a relaxational mode (green shading), $A(\mbox{x})\omega_P(T,\mbox{x})[\omega_P(T,\mbox{x})-i\omega]^{-1}$. (b to d) shows the doping dependence of the intensity of the continuum (b), the relaxational mode (c) and the oscillator (d) above $T_S(\mbox{x})$ and $T_c$. The dashed red line near x$\simeq$0.025 indicates the doping at which the intensity in (b) and (d) becomes negligible. (e) Doping dependence of the mean field transition temperature $\theta(\mbox{x})$.
 \vspace*{0pt}}\label{Fig: RMFit}
\end{figure*}

\begin{figure*}[htpb!]
  \vspace*{-10pt}\centering
  \includegraphics[width=\textwidth]{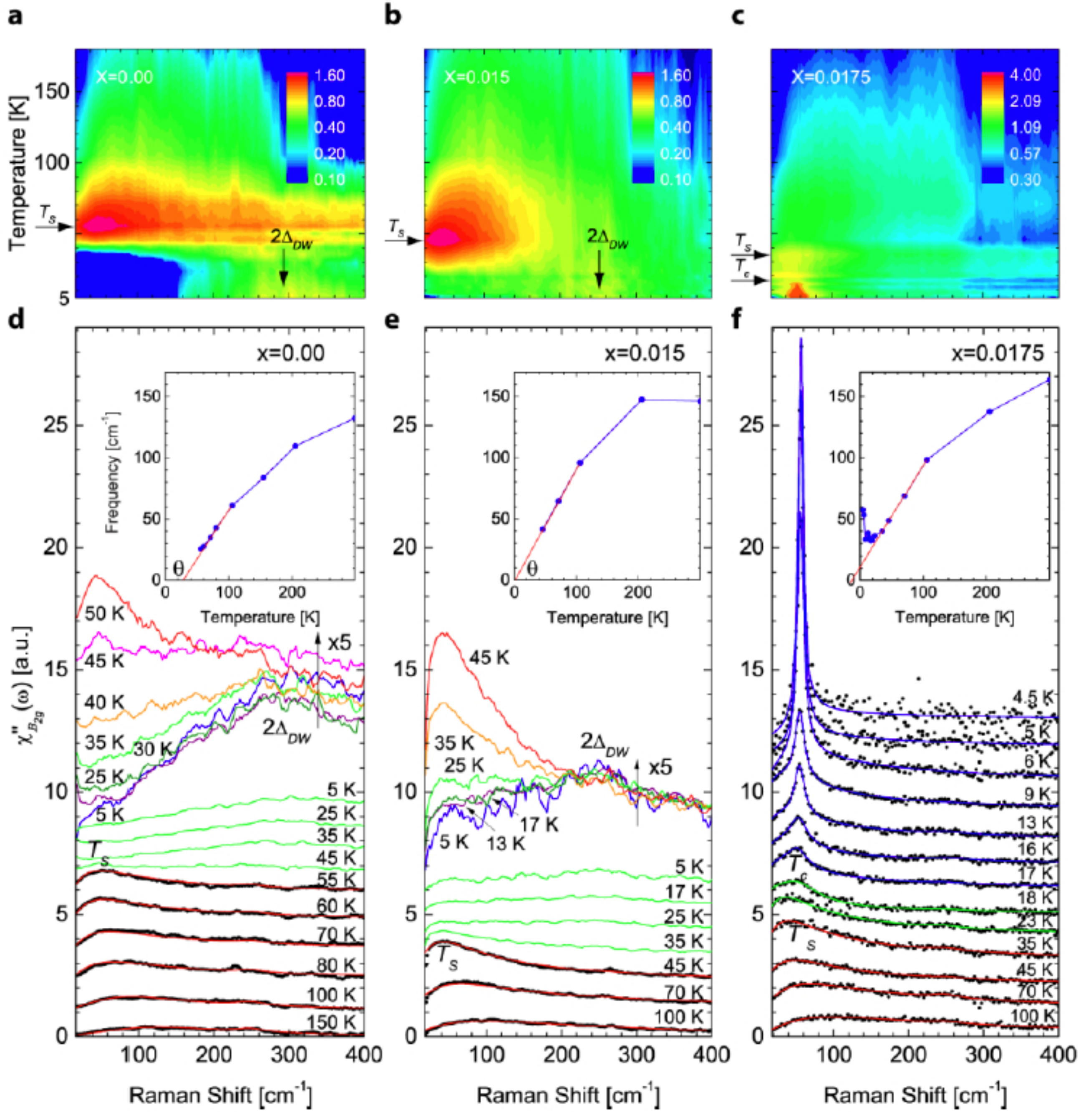}
  \caption{Raman susceptibility $\chi''_{XY}(\omega,T)$ in the $B_{2g}$ symmetry channel at dopings 0$\leq$x$\leq$0.0175. (a to c) Temperature and frequency evolution of $\mbox{Log}(\chi''_{XY}(\omega,T))$ at dopings x=0, x=0.015 and x=0.0175. The structural transition $T_S(\mbox{x})$ is indicated on the temperature axis, and the coherence peak, $2\Delta_{DW}$ on the frequency axis for x=0 and x=0.015. (d to f) $\chi''_{XY}(\omega)$ for $T$$\leq$100~K displaced vertically for clarity. All dopings show the development of the relaxational mode (RM) in the tetragonal phase described by, $A(\mbox{x})\omega_P(T,\mbox{x})[\omega_P(T,\mbox{x})-i\omega]^{-1}$; x=0 and x=0.015 show the development of the coherence peak and spectral weight suppression in the orthorhombic phase; x=0.0175 show the emergence of the sharp resonance in the superconducting phase. The insets display $\omega_P(T,\mbox{x})$ versus temperature.
  \vspace*{0pt}}\label{Fig: RamanData2}
\end{figure*}

\begin{figure*}[!t]
  \vspace*{-10pt}\centering
  \includegraphics[width=\textwidth]{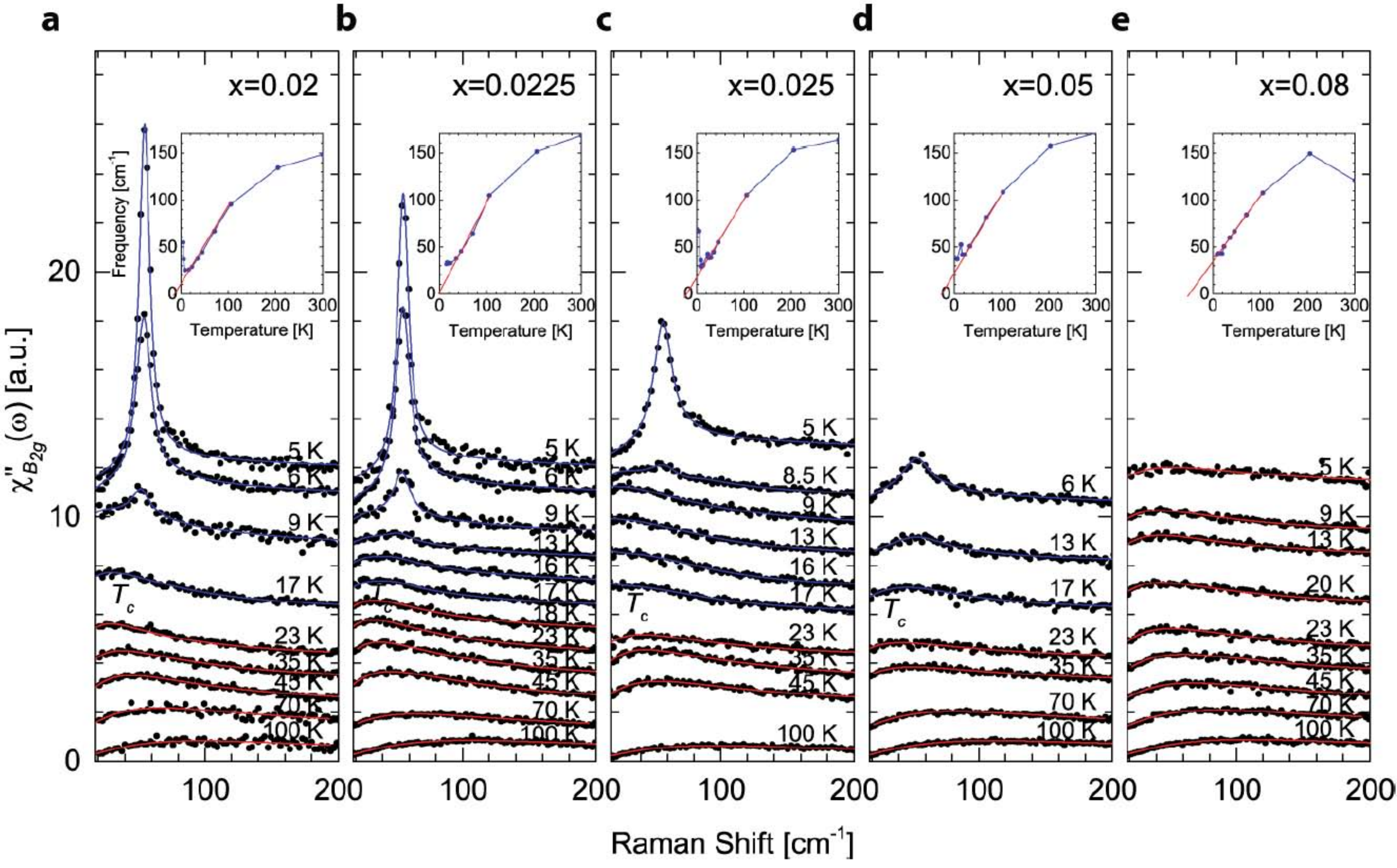}
  \caption{Raman susceptibility $\chi''_{XY}(\omega,T)$ in the $B_{2g}$ symmetry channel at dopings 0.02$\leq$x$\leq$0.08. (a to e) $\chi''_{XY}(\omega,T)$ at various temperatures displaced vertically for clarity. All dopings show the development of the relaxational mode (RM) in the tetragonal phase described by, $A(\mbox{x})\omega_P(T,\mbox{x})[\omega_P(T,\mbox{x})-i\omega]^{-1}$ and the emergence of the sharp resonance in the superconducting phase.
  \vspace*{0pt}}\label{Fig: RamanData3}
\end{figure*}

In the tetragonal phase above $T_S(\mbox{x})$,
$\chi''_{XY}(\omega,T,\mbox{x})$ reveals the emergence of broad
quasielastic scattering (QES) peaked at $\omega_P (T,\mbox{x})$
(Figs.~\ref{Fig: RamanData2},\ref{Fig: RamanData3},\ref{Fig: RMFit}). The
intensity of this feature is weak at high temperatures. Upon cooling,
it softens and gains in intensity, where it reaches a maximum at the
$T_S(\mbox{x})-$line and near $T_c(\mbox{x})$. Below $T_S(\mbox{x})$ the static susceptibility drops rapidly without any observable inflection point at $T_{SDW}(\mbox{x})$, which is congruent with a smoothly developing orthorhombic OP demonstrated in X-ray and neutron diffraction studies \cite{Li2009,Wright2012,Park2012}.


We apply a universal fit to $\chi''_{XY}(\omega,T,\mbox{x})$ with a simultaneous fit of our data as a function of frequency, temperature and doping (See Appendix C, Fig.~\ref{Fig: RMFit2}). Above the $T_S(\mbox{x})$ and $T_c(\mbox{x})$ lines, $\chi''_{XY}(\omega,T,\mbox{x})$ can be decomposed into three components (Fig.~\ref{Fig: RMFit}(a)) which includes a broad QES peak which can be described as a relaxational mode (RM), $\chi^{RM}_{XY}(\omega,T,\mbox{x})\propto A(\mbox{x})[\omega_P(T,\mbox{x})-i\omega]^{-1}$, a continuum and a minor peak at $\simeq$240~cm$^{-1}$. Description of the RM is based on a phenomenological model, Eqs.~\ref{RPA}-\ref{Raman 1} (See Appendix C).
Both the intensity of the continuum and of the $\simeq$240~cm$^{-1}$ mode diminishes rapidly with doping, and vanishes near x$\simeq$0.025
(Figs.~\ref{Fig: RMFit}(c,d)). Importantly, the orbital content of the larger $\gamma$ FS is mainly composed of $d_{xy}$ orbitals, while the $\alpha$ and $\beta$ FSs primarily have $d_{xz}$ and $d_{yz}$ orbital character (Fig.~\ref{Fig: FeAs}(c)) \cite{Zhang2012}. At the M point, the inner (outer) part of the $\delta$/$\epsilon$ FS has $d_{xz}$ and $d_{yz}$ ($d_{xy}$) orbital character. The continuum and the $\simeq$240~cm$^{-1}$ mode likely involve the $\beta$ band as its FS reduces with doping (See Figs.~\ref{Fig: FeAs}(b,d)) with the former due to intraband excitations and the latter due to an interband-like excitation with a 240~cm$^{-1}$ gap consistent with quadrupole excitations as verified by scaling of $\chi^{XY}_0(T,\mbox{x})$ to NQR data (See Section IV). This finding is consistent with first-principle calculations taking into account spin-orbit coupling (Fig.~\ref{Fig: FeAs}(e)).

Figure~\ref{Fig: RMFit}(c) displays the intensity dependence of the RM with doping which is seen to persist for all dopings. Figure~\ref{Fig: RMFit}(e) shows the doping dependence of $\theta(\mbox{x})$ which is observed to decrease close-to linear for increasing dopings becoming negative near x=0.022. This behavior is consistent with that obtained from the analysis of the static Raman susceptibility $\chi_0^{XY}(T,\mbox{x})$ shown in the $T$$-$$\mbox{x}$ phase diagram (Fig.~\ref{Fig: PhaseDiagram}(a)).


The intensity of the RM decreases with doping (Fig.~\ref{Fig: RMFit}(c)). The frequency decreases linearly upon cooling below
$\simeq$100~K for all dopings with the extension crossing
the temperature axis at $\theta(\mbox{x})$ (Insets to Figs.~\ref{Fig: RamanData2}(d-f),\ref{Fig: RamanData3}). The decrease of
$\theta(\mbox{x})$ with doping can be described by a function
$\theta(\mbox{x})=b_1-a_1\mbox{x}$ crossing zero at x=x$_c$$\simeq$0.02 and
becoming negative for x$\gtrsim$0.02 (Fig.~\ref{Fig: RMFit}(e)). In the $T$$-$x phase diagram,
Fig.~\ref{Fig: PhaseDiagram}(a), the $\theta(\mbox{x})-$line is parallel
to the $T_{SDW}(\mbox{x})$ and $T_S(\mbox{x})-$lines \cite{Note1},
approximately 10 and 20~K below, respectively, for x$\lesssim$0.02.


The critical behavior of the susceptibility $\chi_{XY}(\omega,T,\mbox{x})$ manifests in: (1) The enhancement of the static Raman susceptibility $\chi^{XY}_0(T,\mbox{x})$ which scales to the universal response function $[T-\theta(\mbox{x})]^{-1}$ upon cooling for all doping concentrations x with a linear temperature dependence of $\theta(\mbox{x})$; and (2) The gain in intensity and near-linear slowdown of the characteristic fluctuation frequency $\omega_P(T,\mbox{x})\propto T-\theta(\mbox{x})$. The inverse of $\chi_0^{XY}(T,\mbox{x})$, shown in the inset to Fig.~\ref{Fig: PhaseDiagram}(b), exhibits the same linear behavior with temperature as $\omega_P(T,\mbox{x})$ of the RM until $T_S(\mbox{x})$ or $T_c(\mbox{x})$, below which $\chi_0^{XY}(T,\mbox{x})$ rapidly falls off.

\begin{figure*}[htpb]
 \vspace*{-10pt}\centering
 \includegraphics[width=0.7\textwidth]{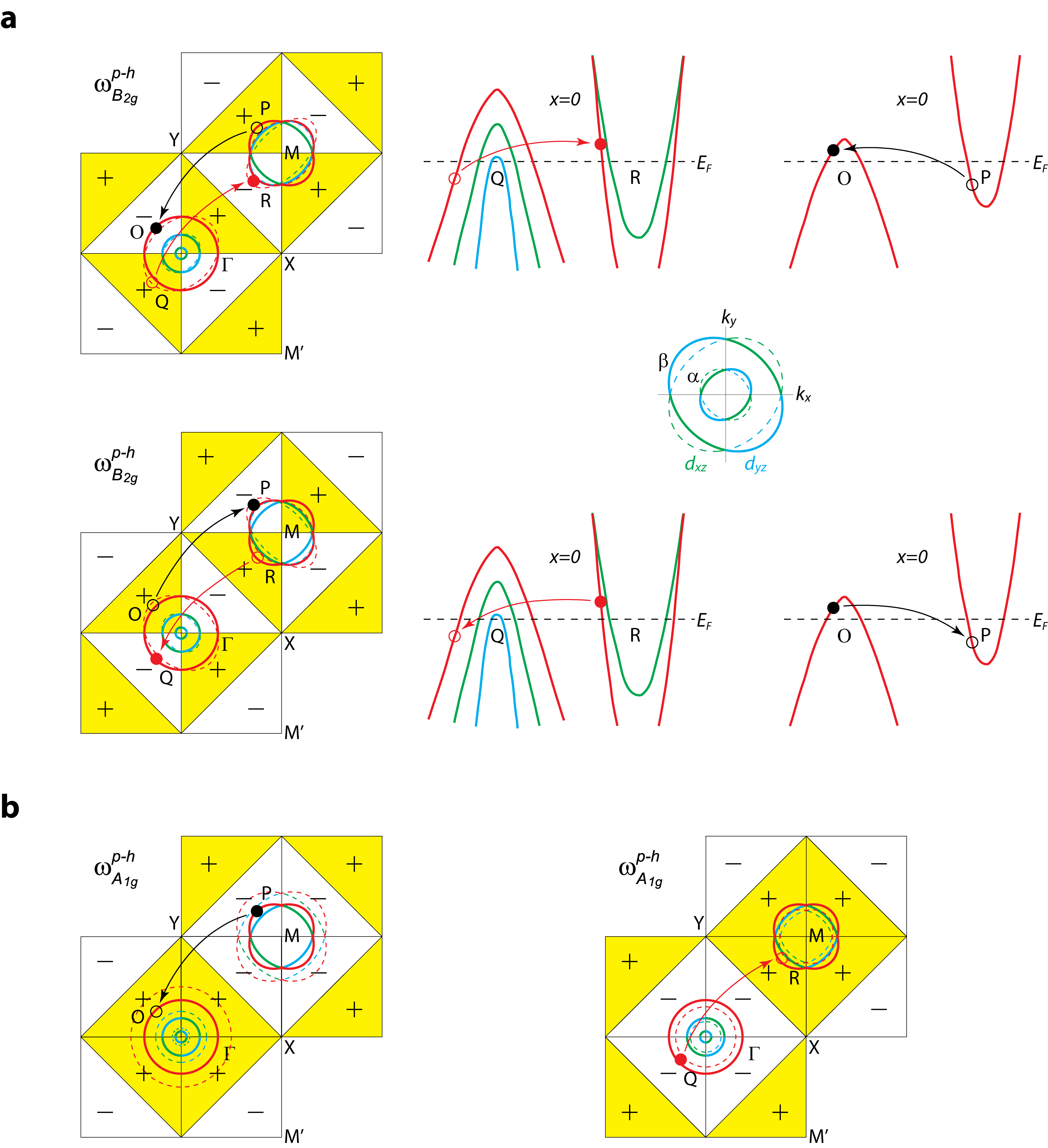}
 \caption{Illustration of symmetry modes in momentum space. (a) Neutral quadrupole charge density modulation with $d^{\pm}$$-$symmetry in the $B_{2g}$$-$symmetry channel. In momentum space the mode is illustrated by positive and negative regions of the BZ. The mode is sustained by $d_{xz}$$-$$d_{yz}$ intra/interband transitions at each the $\Gamma$ and M-point. The coupled interband electron-hole excitations between participating bands at the Γ- and M-point helps to stabilize the $d^{\pm}$$-$mode. (b) $s^{\pm}$ breathing mode in $A_{1g}$ symmetry.\vspace*{0pt}}\label{Fig: SM_QP}
\end{figure*}

Next we reflect on the emergent critical enhancement of $\chi_{XY}(\omega,T,\mbox{x})$ as a result of strong electronic interactions. Potential reasons for the critical behavior include: (1) Electronic coupling to lattice degrees of freedom; (2) Magnetic fluctuations \cite{Kretzschmar2015A,Karahasanovic2015,Chubukov2015} which may invoke the Ising spin-nematic scenario; (3) Charge fluctuations leading to charge order. Here we consider the latter in terms of quadrupole Pomeranchuk fluctuations as the most likely candidate. The partially filled Fe-orbitals with 3$d^6$ configuration give rise to interorbital quadrupole charge fluctuations \cite{Yoshizawa2012,Goto2011,Ma2014}. The critical charge fluctuations in real space are manifested in electron-hole excitations between degenerate $d_{xz}$ and $d_{yz}$ orbitals on Fe-sites which lead to charge transfers as illustrated in Fig.~\ref{Fig: FeAs}(f). This induces a dynamic quadrupole moment of $B_{2g}$ symmetry with nodes along the $X$$-$$Y$ directions (Fig.~\ref{Fig: FeAs}(b)). These are orbital singlet excitations ($\Delta$L=2). In momentum space, the fluctuations lead to dynamic distortions of the FSs around the $\Gamma$- and M-point (Fig.~\ref{Fig: FeAs}(f)) resulting in fluctuating quadrupole moments with nodes along $\Gamma$X and $\Gamma$Y (Fig.~\ref{Fig: SM_QP}(a)). In-phase synchronization of the two FSs leads to $d^{\pm}$ quadrupole deformations which are favored over $d^{++}$ for a dominant repulsive interaction between the $\Gamma$- and M-points (See Appendix E for further details).

These critical quadrupole fluctuations drive the system towards a Pomeranchuk-like instability extended to multibands \cite{Pomeranchuk1959,Hartnoll2014,Oganesyan2001,Edalati2012,Wu2007,Yamase2011}. In a Fermi liquid, the Pomeranchuk instability directly leads to a nematic transition via spontaneous quadrupole deformation of the FSs which freeze with static distortions and in real space the 2-Fe unit cell becomes monoclinic (Fig.~\ref{Fig: FeAs}(g)). The critical behavior of the $B_{2g}$ Raman response foretells the approaching second order Pomeranchuk phase transition at $\theta(\mbox{x})$ which breaks rotational invariance while translational symmetry is preserved. It occurs when the attraction in the $d$-wave channel exceeds a critical threshold \cite{Pomeranchuk1959}.
The Pomeranchuk instability in iron pnictides is special in that it breaks the discrete $C_4$ symmetry via orbital ordering, i.e. a quadrupole lattice in an ordered orbital pattern (See Fig.~\ref{Fig: FeAs}(g)) \cite{Lv2011} but without instigating a density wave (DW) instability.
Similar to Fermi liquids it requires an attraction in the $d$-wave ($B_{2g}$) channel, i.e. an interaction term of the form $g_0 (n_{xz} - n_{yz})^2$ with $g_0<0$ favoring an occupation difference $n_{xz} - n_{yz}$ of the $d_{xz}$ and $d_{yz}$ orbitals. The low-energy anomalies in the $B_{2g}$ Raman data reflects the critical fluctuations associated with the Pomeranchuk instability.
The extraordinary large temperature and frequency range of these fluctuations is consistent with the presence of a QCP defined by a vanishing Weiss-temperature $\theta (\mbox{x})$. The scaling of $\chi^{XY}_0(T,\mbox{x})$ in a two-component fit to NQR data provides compelling evidence of quadrupole-relaxation (See Section IV) \cite{Dioguardi2015}. The range of the critical fluctuations in the $XY$-symmetry channel extend over a much wider temperature range (Figs.~\ref{Fig: PhaseDiagram}(a,b)) than the SDW fluctuations limited to a narrow temperature range above $T_{SDW}(\mbox{x})$ \cite{Steckel2015, Park2012,Wang2012a,Chen2009}.

In conclusion, we find Fe-orbital quadrupole fluctuations display critical behavior foretelling an approaching new ground state below the $\theta$-line Pomeranchuk instability. These results appear to be consistent with other pnictide materials including the 122 and FeSe families suggesting this conclusion may be more universal. In addition, the observation of a similar Weiss-temperature $\theta(\mbox{x})$-like line in the $T$$-$x phase diagram of both Na-111 and Ba-122 systems by Raman \cite{Gallais2013,Gallais2015}
suggests the $\theta$-line to be a universal feature of pnictides. A $\theta$-line is likewise seen in elastic strain measurements \cite{Bohmer2014,Bohmer2015Ar,Yoshizawa2012,Goto2011,Chu2012}. In NMR studies of FeSe, no Curie-Weiss behavior was observed in the relaxation rate $1/T_1T$ above $T_S$ \cite{Bohmer2015a,Baek2015}. In FeSe, NMR is only sensitive to spins since the $^{77}$Se nucleus has spin 1/2 and does not couple to quadrupoles. However, in the 122-family compounds mentioned above which do display Curie-Weiss behavior, the $^{77}$As nucleus has spin 3/2 and does relax into quadrupolar excitations. This observation imply that the Curie-Weiss behavior originates from quadrupoles. The fact that Curie-Weiss behavior is indeed observed in FeSe when using a $C_{66}$ probe \cite{Bohmer2015a} further underlines this conclusion. In Raman studies of Co-doped Ba-122 \cite{Gallais2013,Kretzschmar2015A} (as well as AFe$_2$As$_2$, A=Eu,Er \cite{Zhang2014}), fluctuations were detected over a range from $T_S$ ($\simeq$138~K for x=0) to room temperature.


\section{Density Wave State}

In the orthorhombic phase $\chi''_{XY}(\omega,T,\mbox{x})$ is characterized by a low-frequency suppression of spectral weight and a peak at 2$\Delta_{DW}$ which develops upon cooling observed for x=0 and x=0.015 crystals (Fig.~\ref{Fig: RamanData2}(a,b,d,e)). The peak is at about $\simeq$300~cm$^{-1}$ at low temperatures for x=0. This is near the $\simeq$33~meV gap value reported by STM studies \cite{Zhou2012}. The evolution of the Raman response as a function of frequency and temperature in the low-doping regime is captured in the color contour plots shown in Figs.~\ref{Fig: RamanData2}(a,b). Whether 2$\Delta_{DW}$ starts to develop at $T_S(\mbox{x})$ or at $T_{SDW}(\mbox{x})$ on cooling is obscured by the quasielastic peak (QEP) which rapidly decreases below $T_S(\mbox{x})$ at which point the quadrupole fluctuations freeze due to the broken $C_4$ symmetry. The energy of 2$\Delta_{DW}$ decreases with doping, and for x$\gtrsim$0.0175, this low-frequency suppression and peak are absent.

The 2$\Delta_{DW}$ feature could potentially originate from an SDW gap similar to what has been reported for 122-systems \cite{Zhang2014,Yang2014,Chauviere2011,Nakajima2010,Hu2008}. Here, signatures of the gap develops below $T_{SDW}(\mbox{x})$ which in the 122-family are near-conjoint with $T_S(\mbox{x})$. In NaFe$_{1-x}$Co$_{x}$As, $T_S(\mbox{x})$ and $T_{SDW}(\mbox{x})$ are separated by more than 10~K. In case the 2$\Delta_{DW}$ suppression develops below $T_S(\mbox{x})$, $C_4$ and translational symmetry are broken together and the instability at $T_S(\mbox{x})$ is Kugel-Khomski\v{i}-type or due to quantum mechanical interactions between orbital and spin degrees of freedom as described by Kugel and Khomski\v{i} \cite{Kugel1982}. If the density wave order develops below $T_{SDW}(\mbox{x})$ translational symmetry is first broken at $T_{SDW}(\mbox{x})$.

\begin{figure*}[!t]
  \vspace*{-10pt}\centering
  \includegraphics[width=\textwidth]{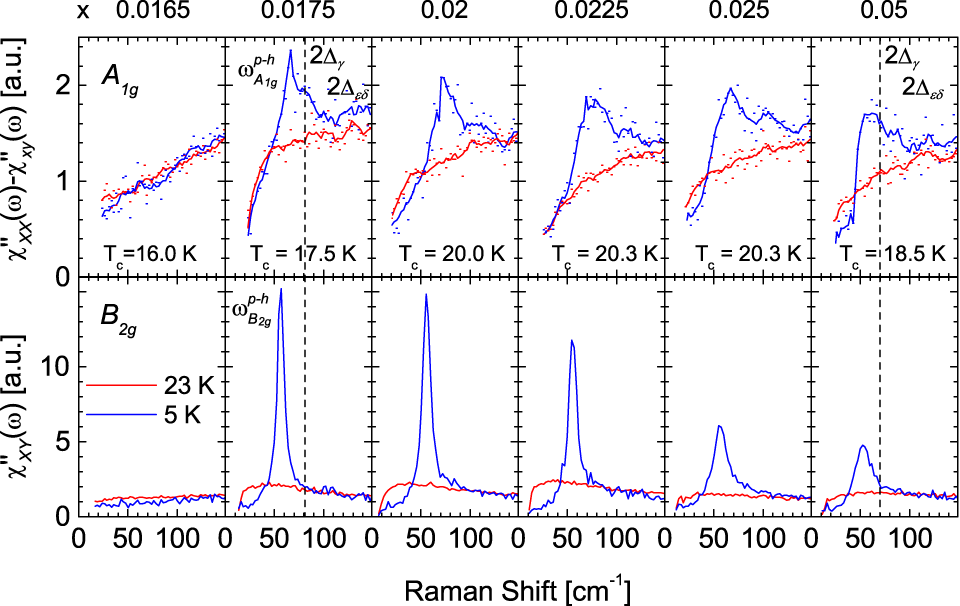}
  \caption{Raman susceptibilities $\chi''_{XX}(\omega)-$$\chi''_{xy}(\omega)$ and $\chi''_{XY}(\omega)$ in the superconducting state. (a) $\chi''_{XX}(\omega)-$$\chi''_{xy}(\omega)$ (top row) and $\chi''_{XY}(\omega)$ (bottom row) in the superconducting (5~K) and normal (23~K) states at doping levels as indicated. The vertical dashed line, shown for x=0.0175 and x=0.05 indicates the lowest superconducting gap determined by ARPES at respectively $\simeq$9~meV \cite{Ge2013} and $\simeq$10~meV \cite{Liu2011}.
  \vspace*{0pt}}\label{Fig: SCModes}
\end{figure*}


\section{Ingap Collective Modes in the Superconducting State}

\begin{figure*}[htpb!]
 \vspace*{-10pt}\centering
 \includegraphics[width=0.8\textwidth]{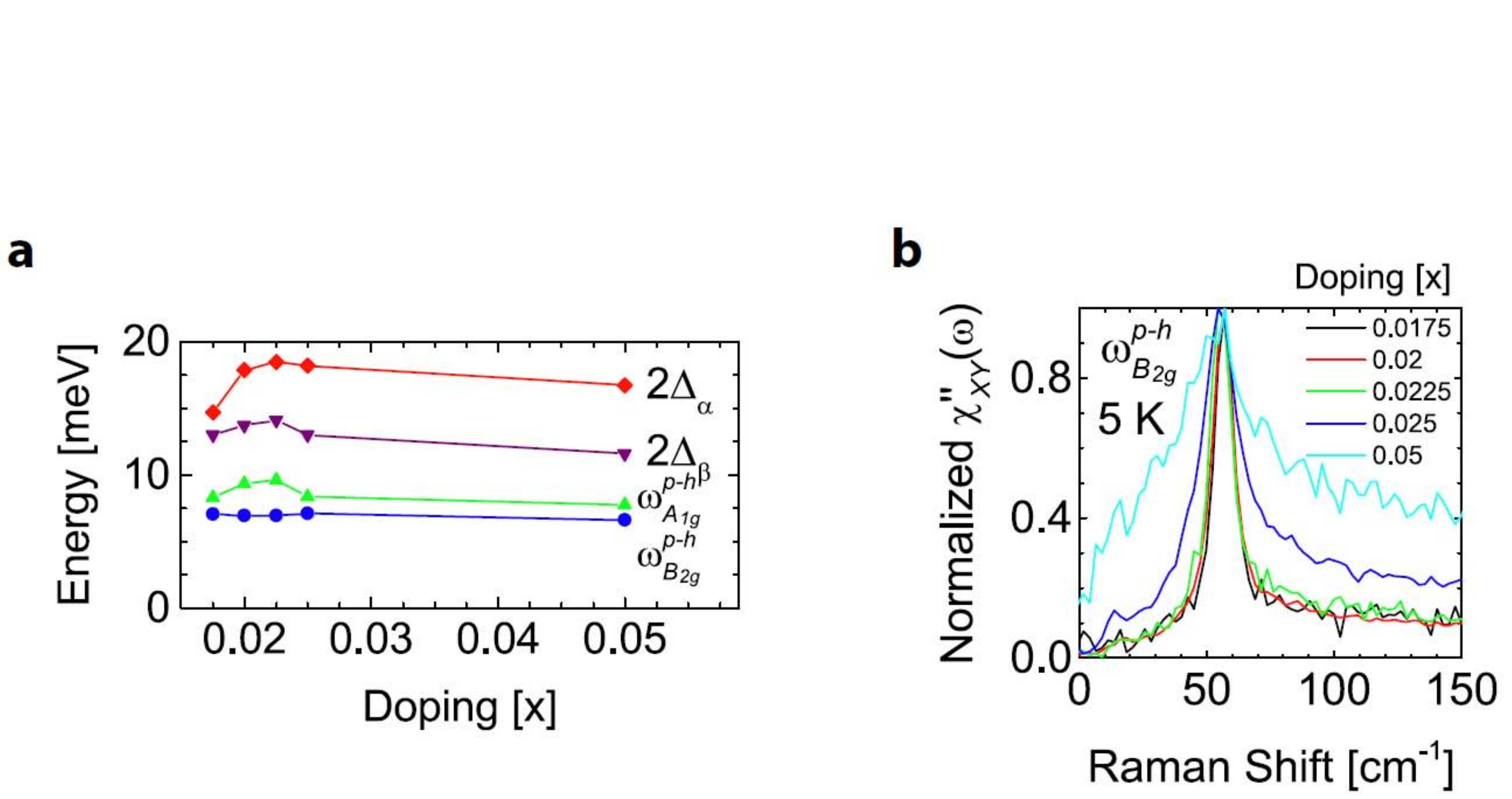}
 \caption{Doping dependence of superconducting gaps, ingap collective modes and their width. (a) Doping dependence of the superconducting gaps and ingap collective modes from Fig.~\ref{Fig: SCModes}. (b) Normalized $\omega_{B_{2g}}$ at various dopings from Fig.~\ref{Fig: SCModes}.\vspace*{0pt}}\label{Fig: SM_Blue}
\end{figure*}

In the superconducting state, for x$\gtrsim$0.0175, the low frequency peak and suppression are absent and $\chi''_{XY}(\omega,T,\mbox{x})$ contains
features in both $A_{1g}$ and $B_{2g}$ symmetry. Below
$T_c(\mbox{x})$, a resonance emerges in $B_{2g}$ symmetry which
sharpens, gains in strength and hardens to
$\omega^{p\mbox{-}h}_{B_{2g}}$$\simeq$7.1~meV upon cooling for
x$\gtrsim$x$_c$ (Figs.~\ref{Fig: RamanData2}(f),\ref{Fig:
RamanData3},\ref{Fig: SCModes}). $\omega^{p\mbox{-}h}_{B_{2g}}$ is the strongest at the lowest temperatures and near x$_c$ at
x=0.0175, and then decreases in strength for increasing doping still
prevailing for x=0.05 and vanishes for x=0.08. The doping dependence
of the superconducting features at 5~K in both
$\chi''_{A_{1g}}(\omega)$ (top row) and $\chi''_{XY}(\omega)$ (bottom
row) in comparison to normal state spectra at 23~K is summarized in
Fig.~\ref{Fig: SCModes}. In the top panel,
$\omega^{p\mbox{-}h}_{A_{1g}}$$\simeq$68~cm$^{-1}$ (8.5~meV),
2$\Delta_{\gamma}$ and 2$\Delta_{\epsilon\delta}$ are present from
x=0.0175 to x=0.05, and are nearly independent of doping (Fig.~\ref{Fig: SM_Blue}(a)).
2$\Delta_{\gamma}$ and 2$\Delta_{\epsilon\delta}$ are consistent with
ARPES \cite{Ge2013,Liu2011} and are assigned as pair-breaking
excitations across the corresponding superconducting gaps
(Fig.~\ref{Fig: SCModes}). The width of
$\omega^{p\mbox{-}h}_{B_{2g}}$ is less than 1~meV for x$\leq$0.0225 (Fig.~\ref{Fig: SM_Blue}(b)) whereafter it broadens and its
intensity diminishes gradually until it vanishes before x=0.08.
$\omega^{p\mbox{-}h}_{A_{1g}}$ and $\omega^{p\mbox{-}h}_{B_{2g}}$ qualify as true ingap excitations as
their energy lies below the minimal quasiparticle gap,
$2\Delta_{\gamma}$ (Fig.~\ref{Fig: SCModes}). In contrast to the superconducting gap feature at the 2$\Delta$ threshold which is characterized by a square-root divergence \cite{Klein1984}, the ingap collective modes appear as sharp delta-function-like resonances.


\begin{figure*}[!t]
  \vspace*{-10pt}\centering
  \includegraphics[width=\textwidth]{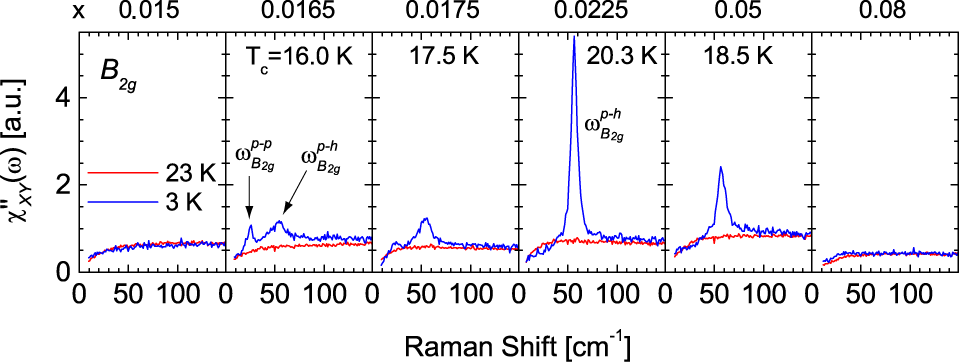}
  \caption{Raman susceptibility $\chi''_{XY}(\omega)$ in the superconducting state. $\chi''_{XY}(\omega)$ in the $B_{2g}$ symmetry channel in the superconducting (3 K) and normal (23 K) state for a laser excitation of 647 nm (1.91 eV) at doping levels as indicated.
  \vspace*{0pt}}\label{Fig: SCModes_Red}
\end{figure*}

Next we interpret the spectrum of collective modes as they may present pertinent information of the superconducting state \cite{Bardasis1961,Klein1984,Chubukov2009,Klein2010,Scalapino2009,Khodas2014,Kretzschmar2013,Bohm2014}. Early studies focused on the Bardasis-Schrieffer mode in BCS single band $s$-wave superconductors where attraction in a non-$s$-wave particle-particle ($p$-$p$) channel would result in the Bardasis-Schrieffer mode forming below the 2$\Delta$ gap edge.\cite{Note2}
In multiband superconductors with weak interband interactions, which applies to MgB$_2$, the Leggett mode results from coherent Cooper pair interband tunneling \cite{Leggett1966,Blumberg2007,Klein2010}. In multiband superconductors with strong interband interactions, which applies to pnictides including NaFe$_{1-x}$Co$_{x}$As, the Leggett mode is pushed above the 2$\Delta$ gap edge where it becomes overdamped and is therefore undetectable.

Recently, Chubukov \textit{et al.} predicted a new ingap exciton in pnictides to appear in $A_{1g}$ symmetry below 2$\Delta$ consistent with a condensate with $s^{\pm}$ symmetry \cite{Chubukov2009,Klein2010}. Rather than Cooper pairs, this mode is composed of particle-hole ($p$-$h$) pairs forming a bound exciton in $A_{1g}$ symmetry. A Raman study of collective modes in multiband superconductors predicted a new $p$-$h$ mode in $B_{2g}$ symmetry below 2$\Delta$ and also discusses the Bardasis-Schrieffer mode \cite{Khodas2014}.

\subsection{Particle-hole Exciton Modes}

\begin{figure*}[htpb]
  \vspace*{-10pt}\centering
  \includegraphics[width=0.75\textwidth]{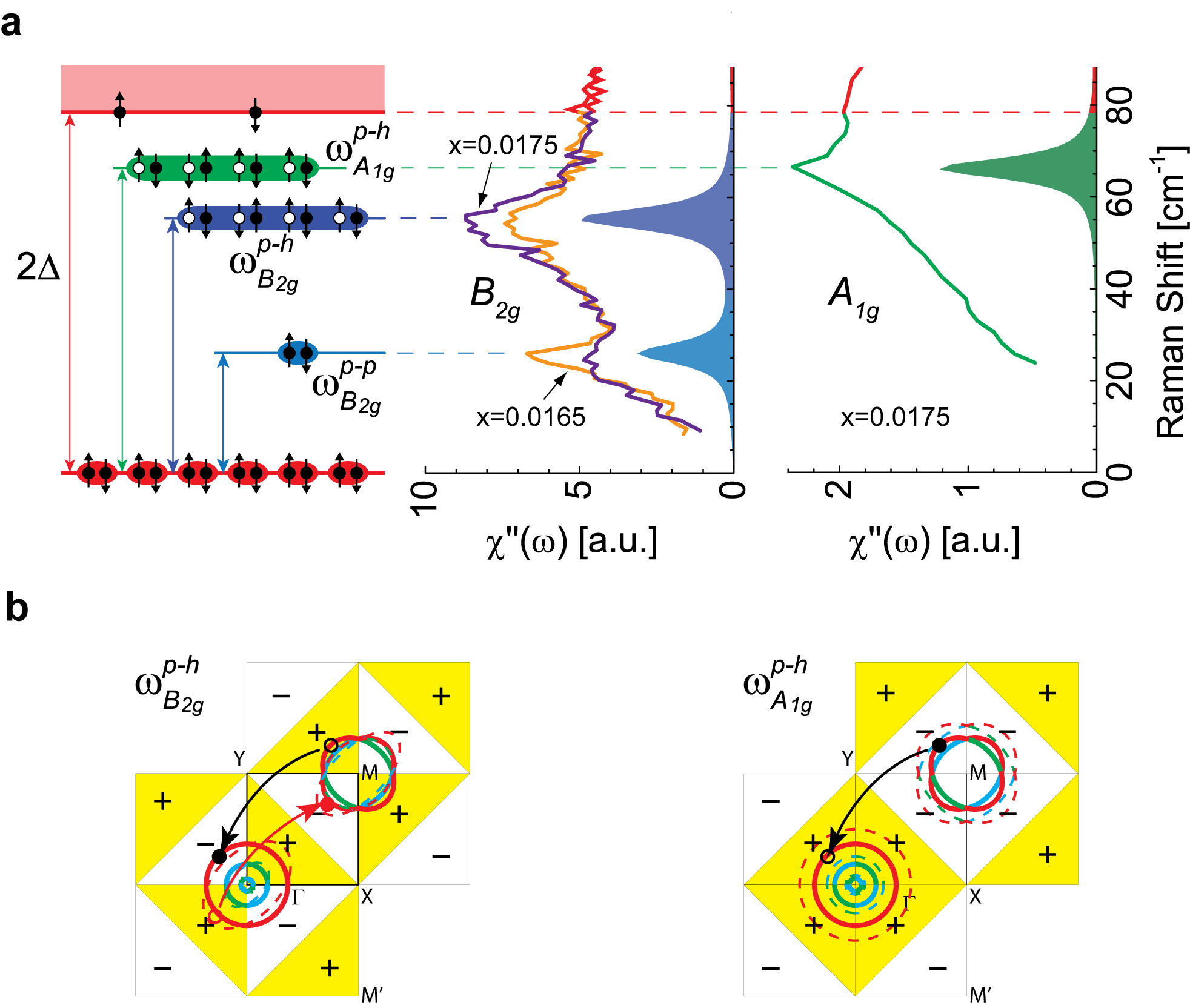}
  \caption{Energy diagram of the superconducting state. (a) Energy diagram of the superconducting state including the superconducting gap 2$\Delta$ and the ingap collective modes $\omega^{p\mbox{-}h}_{A_{1g}}$, $\omega^{p\mbox{-}h}_{B_{2g}}$ and $\omega^{p\mbox{-}p}_{B_{2g}}$ shown together with their spectroscopic signatures in the Raman data. The red horizontal dashed line indicating the lowest superconducting gap coincides with that determined by ARPES at $\simeq$9~meV for x=0.0175 \cite{Ge2013}. The $A_{1g}$ spectrum for x=0.0175 (green) was obtained at 5~K with an excitation energy of $\omega_L$=2.6~eV, and the $B_{2g}$ spectra for x=0.0165 (orange) and x=0.0175 (purple) were obtained at 3~K and $\omega_L$=1.91~eV. The modes determined from a two-band model calculation (not included) are shown together with the $A_{1g}$ and $B_{2g}$ spectra for illustration, and have the area fully colored below the modes of $\omega^{p\mbox{-}h}_{A_{1g}}$ (green), $\omega^{p\mbox{-}h}_{B_{2g}}$ (dark blue) and $\omega^{p\mbox{-}p}_{B_{2g}}$ (light blue).
  (b) Illustration of the symmetry of the BZ for the $\omega^{p\mbox{-}h}_{A_{1g}}$ and $\omega^{p\mbox{-}h}_{B_{2g}}$ modes in the $p$-$h$ channel having respectively $s^{\pm}$ and $d^{\pm}$ symmetry. (See Appendix E).
  \vspace*{0pt}}\label{Fig: CollectiveModes}
\end{figure*}

We assign $\omega^{p\mbox{-}h}_{A_{1g}}$ to the $p$-$h$ charge exciton predicted by Chubukov \textit{et al.} \cite{Chubukov2009,Klein2009}. This $p$-$h$ mode in $A_{1g}$ symmetry is represented by in-phase breathing between the electron and hole FSs (Fig.~\ref{Fig: CollectiveModes}(b) and Appendix E, Fig.~\ref{Fig: SM_QP}(b)). This breathing mode entails periodic charge transfer between the particle and hole pockets.
The out-of-phase breathing of the particle and hole pockets turns
the repulsion into an effective attraction \cite{Chubukov2009}.
The sign flip of the effective interaction is similar to the effective attraction in the Cooper channel for opposite sign of the OP for the particle and hole pockets.
Hence, both $\omega^{p\mbox{-}h}_{A_{1g}}$ and $s^{\pm}$ depend on the inter-pocket interaction winning over intra-pocket repulsion \cite{Klein2010,Chubukov2009}.
$\omega^{p\mbox{-}h}_{A_{1g}}$ signifies attraction in the $s$-wave channel in much the same way as $\omega^{p\mbox{-}h}_{B_{2g}}$ does in the $d$-wave channel.
If strong enough, such attraction may lead to the Pomeranchuk instability in the $A_{1g}$ channel.

The significant intensity of $\omega^{p\mbox{-}h}_{B_{2g}}$ suggests that it couples to light directly, implying that it is the
$d$-wave counterpart of the $\omega^{p\mbox{-}h}_{A_{1g}}$ exciton \cite{Khodas2014}.
Because $\chi_{XY}(\omega,T,\mbox{x})$ is controlled by a large
coupling constant $g$, the
$\omega^{p\mbox{-}h}_{B_{2g}}$ resonance, which is facilitated by a
positive feedback of the superconductivity, emerges from the normal state RM upon
cooling through $T_c$ while retaining its identity as a bound state of $d^{\pm}$ $p$-$h$ oscillations.
Hence, at higher dopings, where the structural transition is suppressed, these $d$-wave Pomeranchuk fluctuations grow strong and below $T_c$ where low-lying excitations are removed, the RM gains coherence and $\omega^{p\mbox{-}h}_{B_{2g}}$ appears as a sharp resonance.

The immediate consequence of attraction in the $B_{2g}$ channel is the ingap resonant modes below the quasi-particle continuum in the superconducting state.
Hence the attraction in the $XY$ channel leads to a sharp resonance below the $p$-$h$ continuum \cite{Lee2013}.
We note that the attraction causing the resonance is operational in the $p$-$h$ channel, while it is well known that the $p$-$h$ and Cooper channel do not have a separate existence and are combined into a single ingap mode \cite{Bardasis1961}.
It was shown that if the superconducting OP changes sign on different sheets of the FS the two channels disentangle \cite{Khodas2014}.
This explains the presence of two rather than one peaks in the underdoped regime.


\subsection{Bardasis-Schrieffer Collective Mode}

For dopings x$\leq$0.0175 and temperatures $\simeq$3~K, a new weak mode appears at $\omega^{p\mbox{-}p}_{B_{2g}}$=$\simeq$25~cm$^{-1}$ (3.1~meV) which becomes stronger for decreasing doping while $\omega^{p\mbox{-}h}_{B_{2g}}$ weakens considerably (Figs.~\ref{Fig: SCModes_Red},\ref{Fig: CollectiveModes}(a)). $\omega^{p\mbox{-}p}_{B_{2g}}$, which we attribute to a Bardasis-Schrieffer mode, exists only in a narrow doping window to the right for both the $T_S(\mbox{x})$ and $T_{SDW}(\mbox{x})$ lines \cite{Steckel2015,Deng2015,Wright2012,Tan2013,Note4}. The maximum $\chi^{XY}_0(T,\mbox{x})$ in Fig.~\ref{Fig: PhaseDiagram}(a) tracks the known part of the $T_S(\mbox{x})$-line and at higher dopings is then observed to slightly curve in towards lower dopings for decreasing temperatures but below $\theta(\mbox{x})$ in a region we will name SC2 (See Figs.~\ref{Fig: PhaseDiagram}(a),\ref{Fig: SCModes_Red},\ref{Fig: CollectiveModes}(a)). In contrast to the detrimental effect of the DW state with the DW gap depleting the density of states, superconductivity below $\theta(\mbox{x})$ is not obstructed by the Pomeranchuk instability. Thus, superconductivity in the orthorhombic phase appears in the narrow doping window below $\theta(\mbox{x})$ (region SC2).

The Bardasis-Schrieffer mode is excited indirectly by photons as the transformation of a $p$-$h$ into a Cooper pair requires assistance of the condensate \cite{Bardasis1961,Monien1990,Khodas2014}.
For $g$$>$0, pairing in the $d$-wave channel provides the conditions for the Bardasis-Schrieffer mode to exist. Figure~\ref{Fig: CollectiveModes}(a) shows an energy diagram of the superconducting state including the superconducting gap 2$\Delta$ and the ingap collective modes $\omega^{p\mbox{-}h}_{A_{1g}}$, $\omega^{p\mbox{-}h}_{B_{2g}}$ and $\omega^{p\mbox{-}p}_{B_{2g}}$ shown together with their spectroscopic signatures in the Raman data.


\section{Quantum Critical Point Inside the Superconducting dome}

Beneath the superconducting dome but above $\theta(\mbox{x})$, which we will name SC1, the susceptibility diverges upon approaching $\theta(\mbox{x})$ and the $\omega^{p\mbox{-}h}_{B_{2g}}$ exciton acquires extraordinary strength. However, in region SC2, below $\theta(\mbox{x})$ the Pomeranchuk fluctuations are gapped and the $\omega^{p\mbox{-}h}_{B_{2g}}$ exciton susceptibility is rapidly suppressed. Upon decreasing the distance to $\theta(\mbox{x})$, i.e. at lower dopings away from x$_c$, the Bardasis-Schrieffer mode gets sharper by borrowing spectral weight from the $\omega^{p\mbox{-}h}_{B_{2g}}$ exciton, (Figs.~\ref{Fig: SCModes_Red},\ref{Fig: CollectiveModes}(a)). The interaction between $\omega^{p\mbox{-}h}_{B_{2g}}$ and $\omega^{p\mbox{-}p}_{B_{2g}}$ versus doping is similar to that discussed for FeSe in Ref.~\onlinecite{Khodas2014}.

The existence of the two superconducting regions SC1 and SC2 which feature the doping-dependent $\omega^{p\mbox{-}h}_{B_{2g}}$ and $\omega^{p\mbox{-}p}_{B_{2g}}$ exciton modes (Fig.~\ref{Fig: SCModes_Red}), separated by $T=\theta(\mbox{x})$, defines a QCP at x$_c$ lying beneath the superconducting dome (Fig.~\ref{Fig: PhaseDiagram}(a)). The location of the boundary between SC1 and SC2 is affected by the competition between the nematic and superconducting orders for carriers \cite{Nandi2010,Moon2012a,Fernandes2012}. Below the Pomeranchuk instability at $\theta(\mbox{x})$, the Pomeranchuk fluctuations vanish and SC2 is characterized by a rhombohedral primitive unit cell, broken $C_4$ symmetry and a quadrupole lattice ordered in an orbital pattern (Fig.~\ref{Fig: FeAs}(g)). In SC1, the critical fluctuations become quantum in nature, and upon decreasing the nonthermal control parameter x from the overdoped regime, $\omega^{p\mbox{-}h}_{B_{2g}}$ gain in strength upon approaching $x_c$. When crossing into SC2, the intensity of the $\omega^{p\mbox{-}h}_{B_{2g}}$ resonance collapses and $\omega^{p\mbox{-}p}_{B_{2g}}$ appears indicative of a QCP occurring at x$_c$. Hence, with doping as a control parameter, we probe spectral weight transfer from the strong $p$-$h$ $B_{2g}$ exciton to the emerging Bardasis-Schrieffer mode and find signatures of a QCP lying beneath the superconducting dome \cite{Khodas2014}.

The QCP is associated with non-Fermi-liquid behavior and occurring at the Pomeranchuk instability becoming quantum at $\theta (\mbox{x=x}_c)$$\equiv$0 suggests it is driven by quadrupole Pomeranchuk fluctuations. The same scenario may prevail in BaFe$_2$(As$_{1-x}$P$_{x}$)$_2$ where a QCP is clearly present below the superconducting dome \cite{Hashimoto2012}, but where a study using NMR, X-rays and neutrons finds no signatures of a QCP \cite{Hu2015}. Thus suggesting the criticality or the QCP arises from the quadrupole Pomeranchuk QCP.

We find that the criticality or the QCP does not arise from either the structural or SDW transitions in support of the quadrupole Pomeranchuk QCP presented in the main text. This conclusion is supported by recent theoretical studies of superconductivity driven by nematic fluctuations at or near a nematic QCP which find that: pairing in the $s$-wave channel is boosted by $d$-wave symmetry fluctuations \cite{Lederer2015}; near a QCP and Pomeranchuk transitions, superconductivity is strongly enhanced \cite{Metlitski2015}. This study concludes that superconductivity is determined by a delicate interplay between the two competing effects, the pairing tendencies of OP fluctuations and strong non-Fermi-liquid effects due to electronic fluctuations; considering a microscopic model, the nematic and SDW transitions merge below a temperature $T_{merge}$$<$$T_c$ and continues to zero temperature as a first-order single nematic-SDW transition line \cite{Fernandes2013c}. This study finds superconductivity to have a strong effect on this quantum phase transition allowing strong fluctuations to exist near it; these transition lines backbend due to superconductivity and there may be a shift of the QCP beneath the superconducting dome \cite{Moon2012b}.

The existence of a QCP has been linked to the occurrence of superconductivity across several classes of unconventional superconductors with a superconducting dome surrounding it in the $T$$-$x phase diagram and with optimal $T_c$ near the QCP. It is by now widely believed that critical quantum fluctuations are important for the superconductivity \cite{Coleman2001,Taillefer2010,Abrahams2011}. These fluctuations enhance interactions and result in an enhancement of electronic correlations upon approaching the QCP 
\cite{Hashimoto2012,Walmsley2013,Shibauchi2014,Nakai2010,Ramshaw2015,Shishido2005,Gegenwart2008,Monthoux2007}.

\section{CONCLUSIONS}

We have studied many-body effects leading to unconventional superconductivity and to competing phases of charge, orbital and spin ordering of the Na-111 family of pnictides containing partially filled 3$d$-orbitals. Using polarization-resolved Raman spectroscopy we find that the inter-orbital attractive interaction, which can be tuned by isovalent Co substitution for Fe, makes the system receptive to the Pomeranchuk-like instability with $d$-wave symmetry and that strong critical fluctuations towards this instability dominate the entire tetragonal phase. In the superconducting phase, these fluctuations acquire coherence and undergo a metamorphosis into ingap collective modes of extraordinary strength. Our finding is an example of non-Fermi-liquid behavior, unconventional superconductivity and electronic ordering emerging from strong multi-polar interactions among 3$d$ electrons, which should be a more generic phenomenon relevant to other compounds containing partially filled $d$ or $f$-orbitals.

\begin{acknowledgements}

We thank A.~V.~Chubukov, P.~Coleman, R.~M.~Fernandes, Y.~Gallais, K.~Haule, T.~T.~Ong, I.~Paul, A.~Sacuto, J.~Schmalian and W.-L.~Zhang for discussions. We extend our special thanks to A.~V.~Chubukov for careful reading of the manuscript and for giving important comments and suggestions. The crystal growth efforts at UTK and Rice were supported by the US DOE, BES, through contract DE-FG02-05ER46202. M.K. thanks the University of Iowa for support and acknowledges support from the Israel Science Foundation (grant No. 1287/15). V.K.T. acknowledges support from NSF through Award DMR-1104884. Research at Rutgers was supported by the US Department of Energy, Office of Basic Energy Sciences through Award DE-SC0005463.

\end{acknowledgements}

\appendix\section{Analysis of Raman Spectra}

\begin{figure}[!t]
 \vspace*{-10pt}\centering
 \includegraphics[width=1.0\columnwidth]{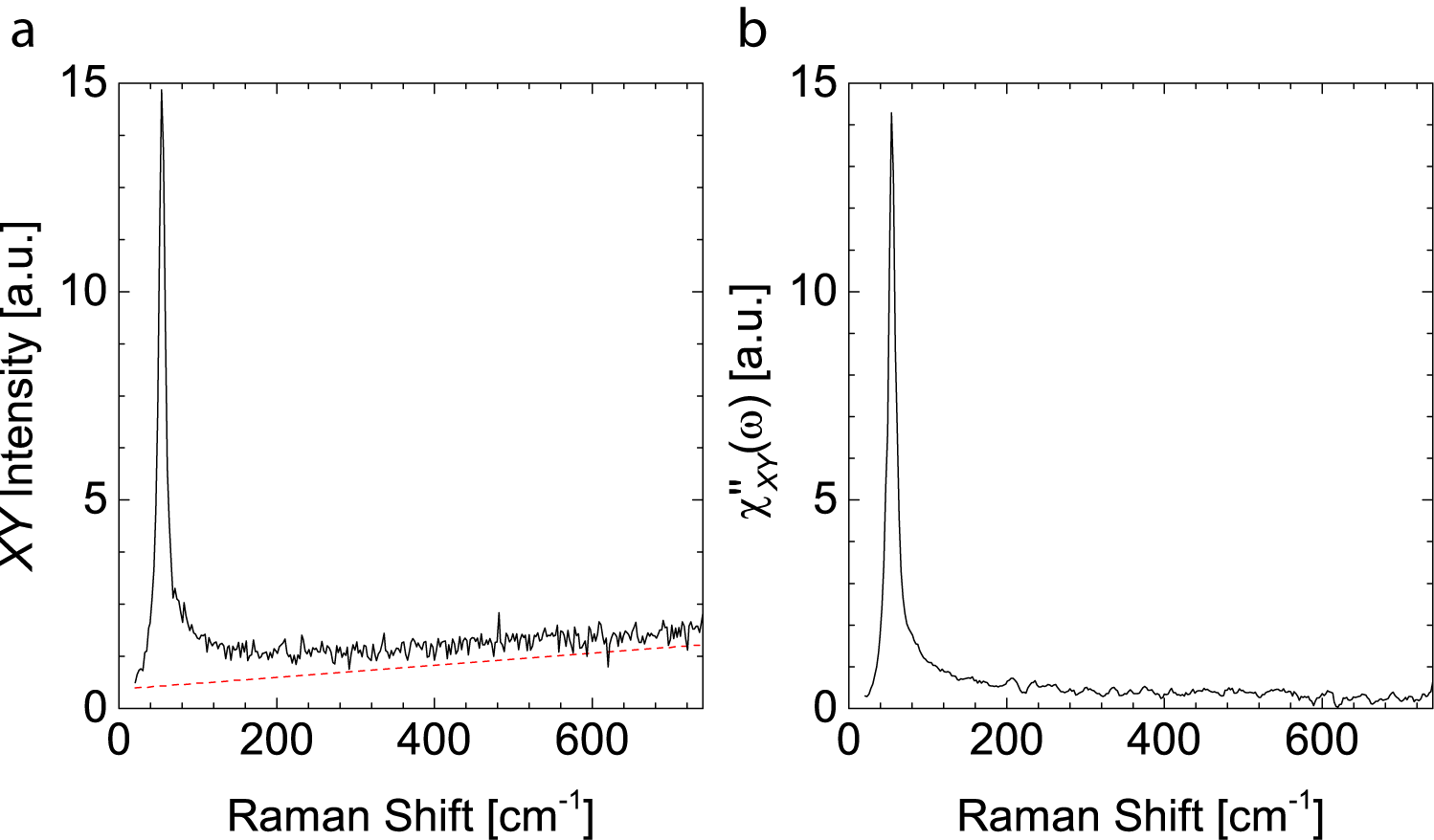}
 \caption{Correction of Raman spectra illustrated for a doping x=0.02 and temperature 5~K in the $XY$ symmetry channel. (a) Raman scattering intensity (black solid line) shown together with a luminescence background $L(\omega)$ (red dashed line) to be subtracted. (b) Raman susceptibility after background subtraction and conversion to $\chi''_{XY}(\omega,T,\mbox{x})$. (See text).
 \vspace*{0pt}}\label{Fig: BG}
\end{figure}

The Raman spectra were corrected for the spectral response of the spectrometer and detector in obtaining the Raman scattering intensity, $I_{e^Ie^S}(\omega) = (1 + n)\chi''(\omega) + L(\omega)$. Here, $L(\omega)$ is a small luminescence background and $\mbox{\bf e}^{I}$ and $\mbox{\bf e}^{S}$ the polarization vectors for the incident and scattered photons for a given scattering geometry with respect to the unit cell (Fig.~\ref{Fig: FeAs}(b)). The recorded Raman intensity was background subtracted with a near-linear line and a constant determined for each polarization geometry as illustrated in Fig.~\ref{Fig: BG}.

\section{Coupling of Pomeranchuk Fluctuations to the Raman Probe}

The goal of this section is to show microscopically that the photons in the $B_{2g}$ configuration are coupled to the local orbital fluctuations shown in Fig.~\ref{Fig: FeAs}(f).
As the orbital character of Raman driven excitations play a central role in our analysis we derive this coupling explicitly.
Apart from the transition between $d_{xz}$ and $d_{yz}$ orbitals the $d_{xy}$ orbital excitations are accessible in the $B_{2g}$ configuration.
The Raman response, Eq.~\ref{response}, is determined by the Raman operators $\widetilde{\rho}^{I,S}$ discussed in details below.

We point out that the standard effective mass approximation \cite{Klein1984,Devereaux2007,Strohm1997} is applicable only to pockets derived from a single non-degenerate band with well defined orbital content.
This is the case for the $\gamma$-pocket derived predominantly from $d_{xy}$ orbitals, see Figs.~\ref{Fig: FeAs}(c,d,e).
Within the effective mass approximation, however to the extent that this pocket is approximately circular the $B_{2g}$ coupling to $\gamma$ pocket is relatively weak, and we focus on the other two electron and hole pockets at the $M$ and at the $\Gamma$ points, respectively.

Consider the Raman coupling to the $\alpha/\beta$ hole pockets first.
It is convenient to use the symmetry constrained $\mathbf{k} \mathbf \cdot \mathbf{p}$ Luttinger Hamiltonian \cite{Cvetkovic2013},
\begin{align}\label{Gamma}
\mathcal{H}^{\Gamma} (\bm{k}) =
\begin{bmatrix}
\epsilon_{\Gamma} + \frac{k^2}{2 m_{\Gamma}} + 2 \tilde{a}  k_x k_y & \tilde{c}(k_x^2 - k_y^2) \\
\tilde{c}(k_x^2 - k_y^2) & \epsilon_{\Gamma} + \frac{k^2}{2 m_{\Gamma}} -  2 \tilde{a} k_x k_y
\end{bmatrix}\, .
\end{align}
Here the parameters $\epsilon_{\Gamma}$, $m_{\Gamma}$, $\tilde{a}$ and $\tilde{c}$ are determined by a fit to the five-band tight-binding model or by first principle calculations which are tabulated in Ref.~\onlinecite{Cvetkovic2013} for selected iron superconductors. We set $\tilde{a}=\tilde{c}$ which corresponds to circular hole FSs.
At the $\Gamma$-point, $\bf{k}=0$, the two Bloch states are degenerate.
These states are characterized by well defined orbital content, and we denote the creation operators of these states by $d^{\dag}_{\alpha =1(2), \bf{k}} = d^{\dag}_{xz(yz),\bf{k}}$.
In terms of these operators the Hamiltonian, \eqref{Gamma} takes the form,
$\hat{\mathcal{H}}^{\Gamma} (\bm{k}) = \sum_{\alpha,\beta=1,2} \mathcal{H}^{\Gamma}_{\alpha,\beta} (\bm{k})  d^{\dag}_{\alpha, \bm{k}} d_{\beta, \bm{k}}$.
The representation \eqref{Gamma} is referred to as orbital to be contrasted with the band representation obtained by diagonalization of \eqref{Gamma}.
The Raman coupling to the $\alpha/\beta$ pockets is a matrix in orbital space \cite{Yamase2013aa,Khodas2015},
\begin{align}\label{Gamma2}
\widetilde{\rho}^{I,S}_{\Gamma}=\sum_{i,j} e_{i}^Ie_{j}^S\sum_{\bm{k}}
\sum_{s,t}
\frac{\partial^2 \mathcal{H}^{\Gamma}_{s,t}(\bm{k}) }{ \partial k_i \partial k_j} d^{\dag}_{s\bm{k}}  d_{t\bm{k}} \, .
\end{align}
In the single-band approximation the orbital indices are redundant, and the more familiar effective mass approximation results.

Substitution of  Eq.~\eqref{Gamma2} in Eq.~\eqref{Gamma} gives for the Raman vertex in $B_{2g}$ geometry
\begin{align}\label{Gamma3}
\widetilde{\rho}^{I,S}_{\Gamma} \propto \sum_{\bm{k}} (d^{\dag}_{xz,\bm{k}}d_{xz,\bm{k}} - d^{\dag}_{yz,\bm{k}}d_{yz,\bm{k}} ).
\end{align}
Therefore, the physical meaning of the $B_{2g}$ Raman probe is the quadrupole excitations causing orbital population imbalance as illustrated in Fig.~\ref{Fig: FeAs}(f).

To understand the implications of the $B_{2g}$ Raman probe in the band representation one diagonalizes the Hamiltonian \eqref{Gamma} which yields $\alpha$ and $\beta$ bands with Bloch states created by the operators $\alpha_{\bf{k}}^{\dag}$ and $\beta_{\bf{k}}^{\dag}$, respectively.
In the band representation the Raman vertex takes the form
\begin{align}\label{Gamma4}
\widetilde{\rho}^{I,S}_{\Gamma} \propto \sum_{\bm{k}} \sin 2 \phi_{\bm{k}} (\alpha^{\dag}_{\bm{k}}\alpha_{\bm{k}} - \beta^{\dag}_{\bf{k}}\beta_{\bm{k}} )
+
\sum_{\bm{k}} \cos 2 \phi_{\bm{k}} (\alpha^{\dag}_{\bm{k}}\beta_{\bm{k}} + \beta^{\dag}_{\bm{k}}\alpha_{\bm{k}} )
\, ,
\end{align}
where $\phi_{\bf{k}}$ is the angle formed by the vector $\bf{k}$ and the $x$-direction in the BZ.
The first intra-band contribution in \eqref{Gamma4} describes the out-of-phase breathing of the $\alpha$ and $\beta$ bands with the amplitude changing as $\sin 2 \phi_{\bf{k}}$, as shown in Fig.~\ref{Fig: FeAs}(f).
The nodes of the intraband $B_{2g}$ excitation are along $k_x$ and $k_y$ as expected.
We conclude, that the Raman response in $B_{2g}$ symmetry couples directly to the Pomeranchuk fluctuations of the FS.
The second, inter-band part of the coupling \eqref{Gamma4} plays a role in temperature and frequency dependence of the $B_{2g}$ response.

The electron pockets coupling to photons can be analyzed along the same lines as is done above for holes using the same effective Hamiltonian approach.
Instead of Eq.~\eqref{Gamma} we have for the electron pockets \cite{Cvetkovic2013},
\begin{align}\label{M1}
h^{\pm}_M(\bm{k})  =
\begin{bmatrix}
\epsilon_1+\frac{\bm{k}^2}{2m_1} \pm a_1 k_x k_y & - i v_{\pm}(\bm{k}) \\
i v_{\pm}(\bm{k}) & \epsilon_3 + \frac{\bm{k}^2}{2 m_3}  \pm a_3 k_x k_y
\end{bmatrix}\, ,
\end{align}
where the upper and lower signs refer to the two electron pockets,
$v_{\pm} \approx v(\pm k_x+k_y)$, $a_{1,3}$, $m_{1,3}$, $\epsilon_{1,3}$, $v$ are parameters to be fixed by matching to the band structure calculations.
The matrix \eqref{M1} acts for the $+(-)$ signs acts in the space of Bloch states that have $xz$, $xy$ ($yz$, $xy$) orbital content.
Again, the electron equivalent of Eq.~\eqref{Gamma2} tells us that the $B_{2g}$
coupling excites the $\pi$ phase shifted breathing of the two electron pockets.
We obtain for an intra-band contribution
\begin{align}\label{M4}
\widetilde{\rho}^{I,S}_{M} \propto \sum_{\bm{k}} F( \phi_{\bm{k}} )(\delta^{\dag}_{\bm{k}}\delta_{\bm{k}} - \epsilon^{\dag}_{\bm{k}}\epsilon_{\bm{k}} )\, .
\end{align}
Eq.~\eqref{M4} shows that photons in the $B_{2g}$ configuration cause the two electron pockets to breathe with a phase difference of $\pi$.

\section{Relaxational Mode Fitting Procedure}

The shape of the Raman response $\chi''_{XY}(\omega,T,\mbox{x})$ with the relaxational mode (RM) and the emergent critical behavior above $T_S(T,\mbox{x})$ can be described by an expression for interacting susceptibilities given by,
\begin{equation}
 \chi_{XY}(\omega,T,\mbox{x})=\lambda^2\frac{\chi^{(0)}_{XY}(\omega,T,\mbox{x})}{1-g\chi^{(0)}_{XY}(\omega,T,\mbox{x})}\, .
\label{RPA}
\end{equation}
Here, $\lambda$ is the coupling of light to the quadrupole charge density fluctuations in $XY$ symmetry (mainly to the $\beta$ band), $\chi^{(0)}_{XY}(\omega,T,\mbox{x})$ is the non-interacting susceptibility and $g$ is the coupling constant.

\begin{figure}[!t]
 \vspace*{-10pt}\centering
 \includegraphics[width=1.0\columnwidth]{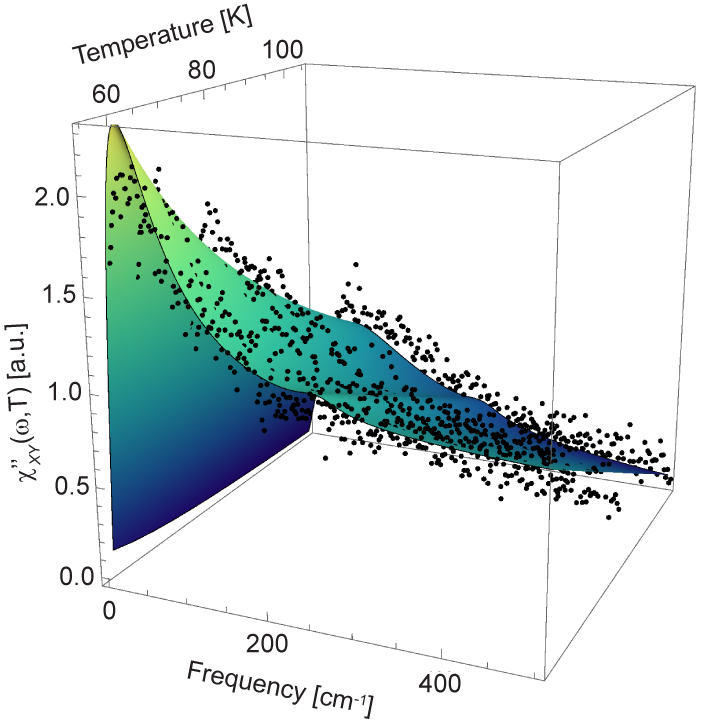}
 \caption{Decomposition of the Raman susceptibility $\chi''_{XY}(\omega,T,\mbox{x})$ in the $B_{2g}$ symmetry channel. $\chi''_{XY}(\omega,T,\mbox{x})$ with the three component fit for x=0 in a range of temperatures above the structural transition showing intensity versus frequency and temperature.
 \vspace*{0pt}}\label{Fig: RMFit2}
\end{figure}

\begin{figure*}[htpb]
 \vspace*{-10pt}\centering
 \includegraphics[width=0.9\textwidth]{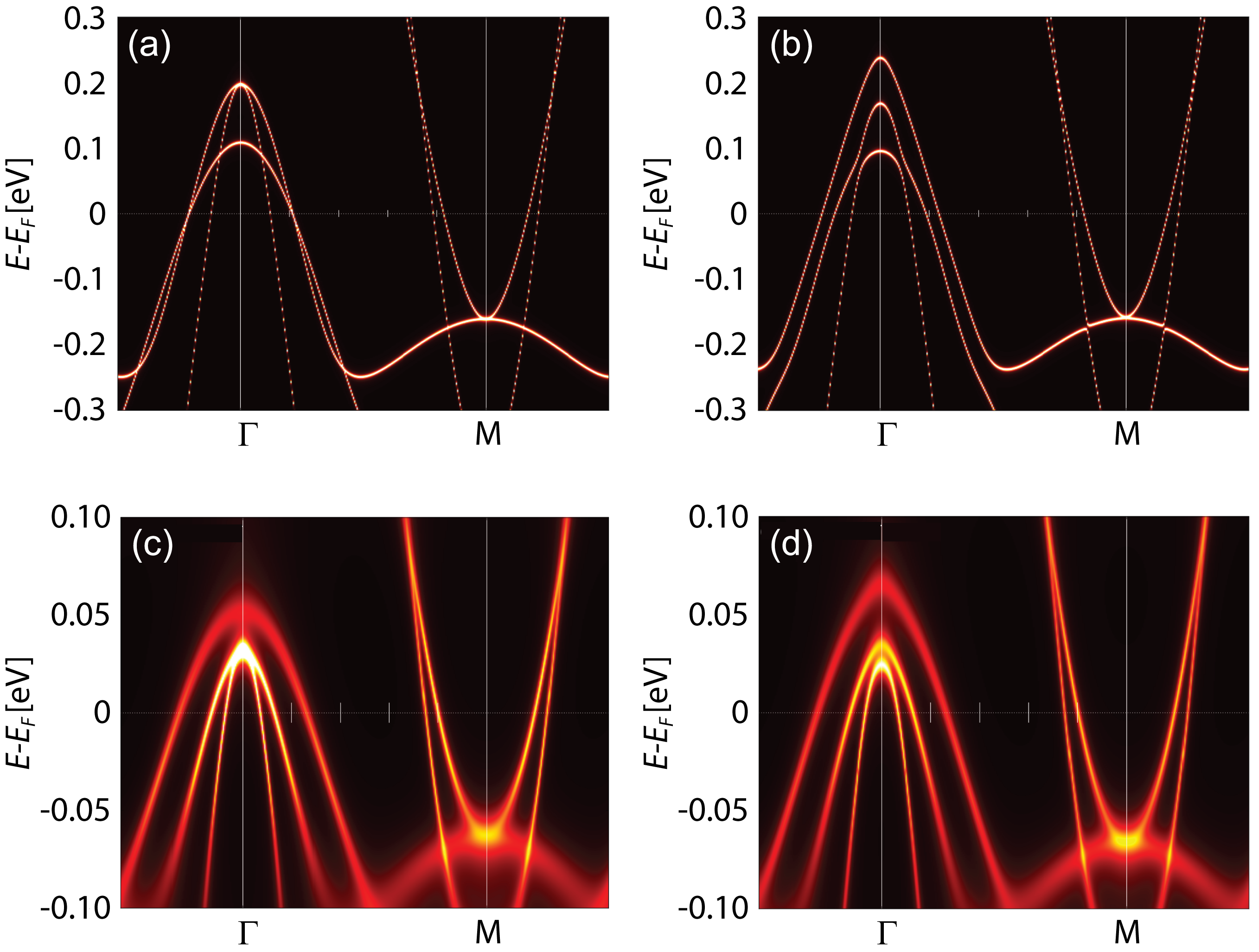}
 \caption{Momentum- and frequency-resolved electronic spectra $A(\mbox{\bf{k}},\omega)$ along $\Gamma-M$ high-symmetry line. (a) DFT without spin-orbit coupling (SOC). (b) DFT with SOC. (c) DFT+DMFT without SOC. (d) DFT+DMFT with SOC. Without SOC, the eigenvalues of the electronic states with $xz$ and $yz$ orbital character are degenerate at the zone center $\Gamma$ point due to the four-fold symmetry of the tetragonal crystal structure in the paramagnetic state. Including SOC lifts this degeneracy, leading to a splitting of the $xz$/$yz$ states, roughly 70~meV at the $\Gamma$ point in the static mean field treatment (DFT+SOC). However, strong electronic correlations strongly renormalize this static mean field splitting to $\simeq$10~meV in the DFT+DMFT+SOC calculation. \vspace*{0pt}}\label{Fig: SOC}
\end{figure*}

Raman spectroscopy as a dynamic probe is well-suited to account for the relaxation processes that determine $[\chi^{(0)}_{XY}(\omega,T,\mbox{x})]''$.
To this end we assume that $[\chi^{(0)}_{XY}(\omega,T,\mbox{x})]''$ is controlled by a single energy scale, $\Gamma_T$.
In addition we assume it to saturate at large frequencies.
Since at low frequencies $[\chi^{(0)}_{XY}(\omega,T,\mbox{x})]'' \propto \omega $ we model it as
$[\chi^{(0)}_{XY}(\omega,T,\mbox{x})]'' = C \mbox{arctan}(\omega/\Gamma_T)$, where $C$ is a constant.
Correspondingly, causality yields
\begin{align}\label{choice_1}
\chi^{(0)}_{XY}(\omega,T,\mbox{x}) = \frac{C}{\pi} \log \frac{(\omega+i\Gamma_T)^2-\Lambda^2}{(\omega+i\Gamma_T)^2},
\end{align}
where $\Lambda$ is a high-frequency cutoff.
The scale $\Gamma_T$ contains contributions from a few elastic and inelastic processes listed below.
While for zero momentum the intraband processes are forbidden \cite{Platzman1965,Yamase2013} this is not so in the present case with finite optical penetration depth.
In addition, the interband transitions at high temperatures make a contribution to
$[\chi^{(0)}_{XY}(\omega,T,\mbox{x})]''$ that scales as $\omega/T$.
Recently, the scattering mechanism that involves both the disorder and long wavelength critical quadrupole fluctuations was shown to give rise to a nearly $T$-linear scattering rate \cite{Hartnoll2014}.
This contribution is expected to grow with doping.
We also note Aslamazov-Larkin corrections due to fluctuations at momentum $Q_{\pi,\pi}$=($\pi/a$,$\pi/a$) \cite{Caprara2005,Kontani2011a,Kontani2014} that are quite different from the quadrupolar fluctuations which also contribute to the scattering rate, $\sim$$2 k_B T$ since the velocities of electrons and holes are antiparallel.
Unlike the above-mentioned contribution, the Aslamazov-Larkin contribution weakens with doping as the deviation from perfect nesting suppresses the coupling of fluctuations at $Q_{\pi,\pi}$ to the zero momentum quadrupolar fluctuations \cite{Paul2014}.
We also note that the elastic scattering off the disorder yields a constant $T$-independent contribution to the scattering rate \cite{Zawadowski1990} which grows with doping.
All of the above scattering processes contribute to $[\chi^{(0)}_{XY}(\omega,T,\mbox{x})]''$.
Our results of the QEP scaling are nevertheless universal because the exact temperature dependence of $\Gamma_T$ at bare level without the effects of quadrupole attraction is not essential.
Essential is that as the relaxation rate in the renormalized Raman susceptibility Eq.~\eqref{RPA} is reduced compared to the bare value $\Gamma_T$.
Therefore, the reduction in the observed width of $[\chi_{XY}(\omega,T,\mbox{x})]''$ reflects the tendency to order at the Pomeranchuk instability.

For $\omega\lesssim\Gamma_T$, substitution of Eq.~\ref{choice_1} in Eq.~\ref{RPA} yields the following relaxational expression for the RM of the $\chi''_{XY}$ susceptibility,
\begin{align}\label{Raman 1}
\chi''_{XY}(\omega,T,\mbox{x}) \propto \frac{\omega\omega_P}{\omega_{P}^2(T,\mbox{x})+\omega^2},
\end{align}
where $\omega_{P}(T,\mbox{x})=\Gamma_T[1/\tilde{g} -\log(\Lambda/\Gamma_T)]$ and $\tilde{g}$=$Cg$. Eq.~\ref{RPA} ensures the critical behavior with temperature above $T_S(\mbox{x})$ and $T_c(\mbox{x})$ of the static susceptibility where $1/\chi_{XY}(0,T,\mbox{x})\propto 1/\chi^{(0)}_{XY}(0,T,\mbox{x})-g$. Here $\theta(\mbox{x})$ is defined by $g$ in terms of $\chi^{(0)}_{XY}(0,\theta,\mbox{x})=1/g$.

Our basic assumption of attraction in the $d^{\pm}$ $p$-$h$
channel follows from the critical enhancement of $\chi_{XY}(\omega,T,\mbox{x})$ and implies $g$$>$0.
For higher dopings, the electron and hole FSs uncouple and our
assumption of attraction in the $d^{\pm}$ $p$-$h$ channel eventually
breaks down.
Hence, the criticality persists but weakens with doping.

We use expression~\ref{Raman 1} with a simultaneous fit of the RM as a function of frequency, temperature and doping dependence. Assuming a weak temperature dependence of the scattering rate we use the expansion $\Gamma_T=\Gamma_{\theta}+\alpha T$. Here $\omega$ and $T$ are fitting variables and $\Gamma_{\theta}$, $\theta$ and $\alpha$ are fitting parameters. Figure~\ref{Fig: RMFit2}(f) shows a universal fit to $\chi''_{XY}(\omega,T,\mbox{x}=0)$ in a range of temperatures above $T_S$ versus frequency and temperature. Above the $T_S(\mbox{x})$ and $T_c(\mbox{x})$ lines, $\chi''_{XY}(\omega,T,\mbox{x})$ can be decomposed into three components (Fig.~\ref{Fig: RMFit}(a)). Both the intensity of the continuum and of the $\simeq$240~cm$^{-1}$
mode diminishes rapidly with doping, and vanishes near x$\simeq$0.025
(Figs.~\ref{Fig: RMFit}(c,d)). Importantly, the orbital content of the larger $\gamma$ FS is mainly composed of $d_{xy}$ orbitals, while the $\alpha$ and $\beta$ FSs primarily have $d_{xz}$ and $d_{yz}$ orbital character (Fig.~\ref{Fig: FeAs}(c)) \cite{Zhang2012}. At the M point, the inner (outer) part of the $\delta$/$\epsilon$ FS has $d_{xz}$ and $d_{yz}$ ($d_{xy}$) orbital character.
The continuum and the $\simeq$240~cm$^{-1}$ mode likely involve the $\beta$ band as its FS reduces
with doping (See Figs.~\ref{Fig: FeAs}(b,d)) with the former due to
intraband excitations and the latter due to an interband-like excitation with a 240~cm$^{-1}$ gap consistent with quadrupole excitations as verified by scaling of $\chi^{XY}_0(T,\mbox{x})$ to NQR data (See Section IV). This finding is consistent with first-principle calculations taking into account spin-orbit coupling (Fig.~\ref{Fig: FeAs}(e)).

Figure~\ref{Fig: RMFit}(c) displays the intensity dependence of the RM with doping which is seen to persist for all dopings. Figure~\ref{Fig: RMFit}(e) shows the doping dependence of $\theta(\mbox{x})$ which is observed to decrease close-to linear for increasing dopings becoming negative near x=0.022. This behavior is consistent with that obtained from the analysis of the static Raman susceptibility $\chi_0^{XY}(T,\mbox{x})$ shown in the $T$$-$$\mbox{x}$ phase diagram (Fig.~\ref{Fig: PhaseDiagram}(a)).

\section{First-principle Band Structure Calculations}

The first-principles calculations use a combination of density functional theory and dynamical mean field theory (DFT+DMFT) \cite{kotliar2006} as in Ref.~\onlinecite{Haule2007}. It is based on the full-potential linear augmented plane wave method implemented in Wien2K \cite{Blaha2001} for carrying out first-principle calculations. The electronic charge is computed self-consistently in the DFT+DMFT density matrix. The continuous time quantum Monte Carlo method \cite{Haule2007,Werner2006} was used to solve the quantum impurity problem using the Coulomb repulsion in its fully rotational form.

We used the experimentally determined lattice structure for NaFeAs with the lattice constants $a=3.94729$~\AA, $b=6.99112$~\AA, and atomic positions $z=0.5$ for Fe, $z_1=0.70234$ for As, $z_2=0.14673$ for Na \cite{Parker2009}. The calculations were done in the paramagnetic state with Coulomb interactions $U$=5.0 eV and $J_H$=0.8~eV at a temperature of $T$=116~K.

\section{Symmetry Modes in Momentum Space}

In $XY$ symmetry, modulations of the FSs around the $\Gamma$- and M-point with nodes along $\Gamma$X and $\Gamma$Y lead to a neutral quadrupole charge density mode with either $d^{++}$$-$ or $d^{\pm}$$-$symmetry of the BZ, Fig.~\ref{Fig: SM_QP}(a). The critical charge fluctuations above the structural transition $T_S$ originate from local electron-hole excitations with charge transfer between the $d_{xz}$ and $d_{yz}$ orbitals on the Fe-sites which introduces a quadrupole moment in $B_{2g}$ or $XY$ symmetry. These local $d_{xz}$ and $d_{yz}$ charge transfer processes are the primary excitations sustaining the quadrupole pattern. Secondly, the phasing between the $\Gamma$- and M-point is dictated by interband interactions across the Fermi level at the $\Gamma$ and M-point as illustrated in Fig.~\ref{Fig: SM_QP}(a). The FSs elongate and squeeze along $\Gamma$M and $\Gamma$M' into quadrants ($d^{++}$) or half quadrants ($d^{\pm}$) defined by the nodes. The quadrupole mode is neutral where the charge at each of the $\Gamma$- and M-point are conserved as well as the overall charge of all participating FSs. Hence, the quadrupole mode results from deformations of the FSs in which charge redistribution by intra and interband transitions causes a quadrupole pattern of changing positive and negative half-quadrant regions of more or less charge. Out-of-phase charge modulation yields $d^{\pm}$$-$symmetry where the FSs at the $\Gamma$- and M-point elongate and squeeze in-phase, while in-phase charge modulation corresponds to $d^{++}$-symmetry. Simultaneous $(\pi,\pi)$ and $(-\pi,-\pi)$ two-electron-hole exchange interactions will promote $d^{\pm}$$-$symmetry.

The in-phase breathing mode predicted by Chubukov \textit{et al.} \cite{Chubukov2009} and later verified by Klein \textit{et al.} \cite{Klein2009} in $A_{1g}$ symmetry is illustrated in Fig.~\ref{Fig: SM_QP}(b). This in-phase breathing mode is a particle-hole exciton which forms in $A_{1g}$ symmetry and is consistent with an $s^{\pm}$ condensate. It is represented by in-phase breathing of the electron and hole FSs and entails charge transfer between the two pockets.

\end{document}